\documentclass[aps,prb,twocolumn,superscriptaddress,showpacs]{revtex4-1}
\usepackage{graphicx}
\usepackage{dcolumn} 
\usepackage{bm}      
\usepackage{color}
\usepackage{multirow}
\usepackage{amsmath,amssymb}
\usepackage{enumitem}
\usepackage{epstopdf}
\usepackage{latexsym}
\usepackage{subfigure}
\usepackage{dsfont} 
\usepackage{wasysym} 

\usepackage{times}
\usepackage{hyperref} 
\hypersetup{
	colorlinks   = true,
	citecolor    = blue,
	linkcolor    = blue,
	urlcolor     = blue
}

\begin{document}

\title{Structural and magnetic properties of two branches of the  Tripod Kagome Lattice family     \\ A$_{2}$RE$_{3}$Sb$_{3}$O$_{14}$ (A = Mg, Zn; RE = Pr, Nd, Gd, Tb, Dy, Ho, Er, Yb)}

\author{Z.~L.~Dun}
\affiliation{Department of Physics and Astronomy, University of Tennessee, Knoxville, Tennessee 37996-1200, USA}

\author{J. Trinh}
\affiliation{Department of Physics, University of California, Santa Cruz, CA 95064, USA}

\author{M.~Lee}
\affiliation{Department of Physics, Florida State University, Tallahassee, FL 32306-3016, USA}
\affiliation{National High Magnetic Field Laboratory, Florida State University, Tallahassee, FL 32310-3706, USA}

\author{E.~S.~Choi}
\affiliation{National High Magnetic Field Laboratory, Florida State University, Tallahassee, FL 32310-3706, USA}

\author{K. Li}
\affiliation{Beijing National Laboratory for Molecular Sciences, State Key Laboratory of Rare Earth Materials Chemistry and Applications, College of Chemistry and Molecular Engineering, Peking University, Beijing 100871, PR China}
\affiliation{Center for High Pressure Science and Technology Advanced Research, Beijing, 100094, PR China}

\author{Y. F. Hu}
\affiliation{Beijing National Laboratory for Molecular Sciences, State Key Laboratory of Rare Earth Materials Chemistry and Applications, College of Chemistry and Molecular Engineering, Peking University, Beijing 100871, PR China}

\author{Y. X. Wang}
\affiliation{Beijing National Laboratory for Molecular Sciences, State Key Laboratory of Rare Earth Materials Chemistry and Applications, College of Chemistry and Molecular Engineering, Peking University, Beijing 100871, PR China}

\author{N. Blanc}
\affiliation{Department of Physics, University of California, Santa Cruz, CA 95064, USA}

\author{A. P. Ramirez}
\affiliation{Department of Physics, University of California, Santa Cruz, CA 95064, USA}

\author{H.~D.~Zhou}
\affiliation{Department of Physics and Astronomy, University of Tennessee, Knoxville, Tennessee 37996-1200, USA}
\affiliation{National High Magnetic Field Laboratory, Florida State University, Tallahassee, FL 32310-3706, USA}

\date{\today}

\begin{abstract}
We present a systematic study of the structural and magnetic properties of two branches of the rare earth Tripod Kagome Lattice (TKL) family A$_{2}$RE$_{3}$Sb$_{3}$O$_{14}$ (A = Mg, Zn; RE = Pr, Nd, Gd, Tb, Dy, Ho, Er, Yb; here, we use abbreviation \textit{A-RE}, as in \textit{MgPr} for Mg$_{2}$Pr$_{3}$Sb$_{3}$O$_{14}$), which complements our previously reported work on \textit{MgDy}, \textit{MgGd}, and \textit{MgEr} \cite{TKL}. The present susceptibility ($\chi_{dc}$, $\chi_{ac}$) and specific heat measurements reveal various magnetic ground states, including the non-magnetic singlet state for \textit{MgPr}, \textit{ZnPr}; long range orderings (LROs) for \textit{MgGd}, \textit{ZnGd}, \textit{MgNd}, \textit{ZnNd}, and \textit{MgYb}; a long range magnetic charge ordered state for \textit{MgDy}, \textit{ZnDy}, and potentially for \textit{MgHo}; possible spin glass states for  \textit{ZnEr}, \textit{ZnHo};  the absence of spin ordering down to 80 mK for \textit{MgEr}, \textit{MgTb}, \textit{ZnTb}, and \textit{ZnYb} compounds. 
The ground states observed here bear both similarities as well as striking differences from the states found in the parent pyrochlore systems. In particular, while the TKLs display a greater tendency towards LRO, the lack of LRO in \textit{MgHo}, \textit{MgTb} and \textit{ZnTb} can be viewed from the standpoint of a balance among spin-spin interactions, anisotropies and non-Kramers nature of single ion state.  While substituting Zn for Mg changes the chemical pressure, and subtly modifies the interaction energies for compounds with larger RE ions, this substitution introduces structural disorder and modifies the ground states for compounds with smaller RE ions (Ho, Er, Yb).
\end{abstract}

\maketitle

\section{Introduction}
The kagome lattice is comprised of corner-sharing triangles in two-dimensions (Fig. 1(a)). Its unique geometry, which combines low dimensionality, low connectivity, and high geometrical frustration, makes it an ideal lattice for realizing exotic states. For example,  the quantum spin liquid (QSL) state \cite{NatureLB} has been proposed for several kagome lattice materials with either Heisenberg spins for  herbertsmithite ZnCu$_3$(OH)$_6$Cl$_2$ \cite{herbertsmithite2005, Nature2012},  polymorph Kapellasite \cite{Kapellasite}, and Zn-doped barlowite \cite{kagomenew}; or with coplanar spins for langasites \cite{Lang1, Lang2}. For most of these systems, lattice distortion or structural disorder \cite{Sitedisorder} complicates an interpretation of the  intrinsic kagome lattice physics, whose effects are largely omitted by theoretical approaches. Importantly, theories of QSLs predict a wide variety of states \cite{RVB, Wen, Wen2007, Z2QSL}, suggesting that tunability of materials parameters will be determinative.  The systems mentioned above represent singular points in a broader phase space of possible ground states, which limits the accuracy of interpretation. It is therefore essential to explore new kagome lattice materials with tunable material parameters and little disorder.

Recently, a new family of compounds, A$_{2}$RE$_{3}$Sb$_{3}$O$_{14}$ (A = Mg, Zn, Co, Mn; RE = rare earth elements) \cite{JSSC2014Co, JSSC2014Mn, TKL, CavaZn, CavaMg}, has been discovered which features an ideal rare earth kagome lattice,  namely,  the ``Tripod Kagome Lattice" (TKL) \cite{TKL}. Below, we use the abbreviated name, \textit{A-RE}, for the TKLs, such as \textit{MgPr} for Mg$_{2}$Pr$_{3}$Sb$_{3}$O$_{14}$. The TKL is a variant of the pyrochlore lattice \cite{RevJG} through partial ion substitution, for which the triangular layers in the pyrochlore lattice along the [111] axis are substituted by A$^{2+}$ ions, resulting in 2D RE$^{3+}$ kagome layers that are separated by A$^{2+}$ triangular layers. In this new structure, the nearest neighbor RE-RE distance within each kagome layer ($\sim$ 3.7 \AA) is much smaller than that of the distance between layers ($\sim$ 6.2 \AA). Given the good interlayer separation, it is clear that for TKLs with a non-magnetic A site (A = Mg, Zn), the dominant magnetic interaction is in the layer, enabling the study of pure kagome physics.

While the two-dimensionality of the TKL lattice controls the interactions, the single ion anisotropy, which is vestigial from the parent pyrochlore structure, gives rise to different spin types (Ising, Heisenberg, and $XY$) for different RE ions with the Ising or the $XY$-spin normal vectors that are neither uniaxial nor uniplanar \cite{TKL}. This particular situation of three distinct tripod-like axes distinguishes the TKL from other kagome lattice materials that possess either Heisenberg or coplanar spins, and potentially gives rise to different states \cite{SOL, KSI, Dipoles, PRBKT, ECO2, ECO1}.  Indeed, various such states have already been revealed by recent studies of four compounds of the TKL family: (i) With Heisenberg spins in \textit{MgGd}, the system shows a long-range ordering (LRO) at 1.65 K, which is consistent with a coplanar 120 $^{\circ}$ $k$ = 0 spin structure where all the spins are in the kagome plane and perpendicular to the local Ising axes. It provides an example of dipolar interaction mandated spin ordering in a 2D system \cite{TKL}. (ii) With  Ising spins in \textit{MgDy}, a phase transition  at $T$ $\sim$ 0.3 K \cite{DyECO} ($T$ = 0.37 K in Ref. \cite{TKL} ) could be related to an ordering among emergent magnetic charge degrees of freedom while a fraction of spin moments remain disordered. Such emergent charge order (ECO) has been proposed theoretically \cite{ECO2, ECO1}, but has never before been experimentally realized. (iii) For \textit{MgNd}, LRO with a non-coplanar $k$ = 0 all-in-all-out state is observed at $T$ = 0.56 K, which may be stabilized by Dzyaloshinskii-Moriya interactions \cite{NdTKL}. (iv) Two transitions at 2.1 K and 80 mK are observed in \textit{MgEr} \cite{TKL} where local $XY$ spins are preserved. The first transition is possibly related to a Kosterlitz-Thouless (KT) vortex unbinding transition \cite{KT}. 

Besides the four examples mentioned above, other members of the \textit{MgRE} family, such as RE = Pr, Tb, Ho, Yb, could prove interesting. In the parent pyrochlore compounds with these four RE elements, different ground states have been reported, including a QSL \cite{TbTi}, a dipolar spin ice \cite{HoTi}, exchange spin ices \cite{PrSn,PrZr} and a quantum spin ice \cite{YbTi1,YbTi2}. What states will occur  when confined to two-dimensions with one of the frustrating spins removed? Moreover, what are the low temperature magnetic properties of the same systems, with marginally larger in-plane lattice constants, such as is realized with \textit{ZnRE}? In this TKL branch, the larger Zn$^{2+}$ ion is expected to modify the spin-spin interaction energies, akin to applying ``chemical pressure" and thus allows rigorous theoretical tests of the low energy phases. Such chemical pressure has been proved to be crucial for selecting ground states in Er and Yb pyrochlores \cite{YbGe, ErGe}. 
The questions raised above provide ample motivation for a systematic exploration of the large range of compounds based on the TKL structure. 

In this manuscript, we synthesized and studied the structural and magnetic properties of two branches of the rare earth TKL family A$_{2}$RE$_{3}$Sb$_{3}$O$_{14}$ (A = Mg, Zn; RE = Pr, Nd, Gd, Tb, Dy, Ho, Er, Yb)  with non-magnetic A site. By combining the experimental probes of X-ray diffraction (XRD), dc and ac susceptibility ($\chi_{dc}$, $\chi_{ac}$), and specific heat ($C(T)$), we show various magnetic ground states for TKLs with different RE ions. By replacing Mg with Zn, we show that the chemical pressure has a small effect on the magnetic properties for TKLs with larger RE ions. We do observe, however, a seeming  increase of site-disorder for TKLs with smaller RE ions by Zn replacement, leading to dramatic changes of ground states compared to those in the Mg branch.

\section{Experimental details}
Polycrystalline samples of A$_{2}$RE$_{3}$Sb$_{3}$O$_{14}$ (A = Mg, Zn; RE = Pr, Nd, Gd, Tb, Dy, Ho, Er, Yb) were synthesized by solid state reactions from powder of RE$_2$O$_3$ (RE = Nd, Gd, Dy, Ho, Er, Yb) / Pr$_6$O$_{11}$ / Tb$_4$O$_7$,  Sb$_2$O$_3$, and MgO/ZnO. For the Zn$_{2}$RE$_{3}$Sb$_{3}$O$_{14}$ family, stoichiometric mixtures were carefully ground and reacted at a temperature of 1200 $^{\circ}$C in air for 3 days with several intermediate grindings, in a manner described previously \cite{CavaZn}. For the Mg$_{2}$RE$_{3}$Sb$_{3}$O$_{14}$ family, a higher reaction temperature of 1300 - 1350 $^{\circ}$C is required to obtain pure TKL phases \cite{TKL}. The room temperature XRD patterns were measured with a HUBER X-ray powder diffractometer with the structural refinements performed using software package \textit{Fullprof-suite}. The $\chi_{dc}$  measurements were performed using a commercial superconducting quantum interference device (SQUID) magnetometer with a magnetic field of 0.05 T. The $\chi_{ac}$ was measured at the National High Magnetic Field Laboratory using the conventional mutual inductance technique at frequencies between 80 Hz and 1000 Hz. The low temperature $C(T)$ measurements were performed in a He3-He4 dilution refrigerator using the semi-adiabatic heat pulse technique. The powder samples were cold-sintered with Ag powder, the contribution of which was measured separately and subtracted from the data.  For all the $C(T)$ data shown below, the magnetic contribution (C$_{mag} (T)$) was obtained by subtracting a lattice contribution estimated from the results of a separate measurement of the non-magnetic isomorph Zn$_2$La$_3$Sb$_3$O$_{14}$ \cite{TKL}.

\begin{figure}[tbp]
	\linespread{1}
	\par
	\begin{center}
		\includegraphics[width= \columnwidth ]{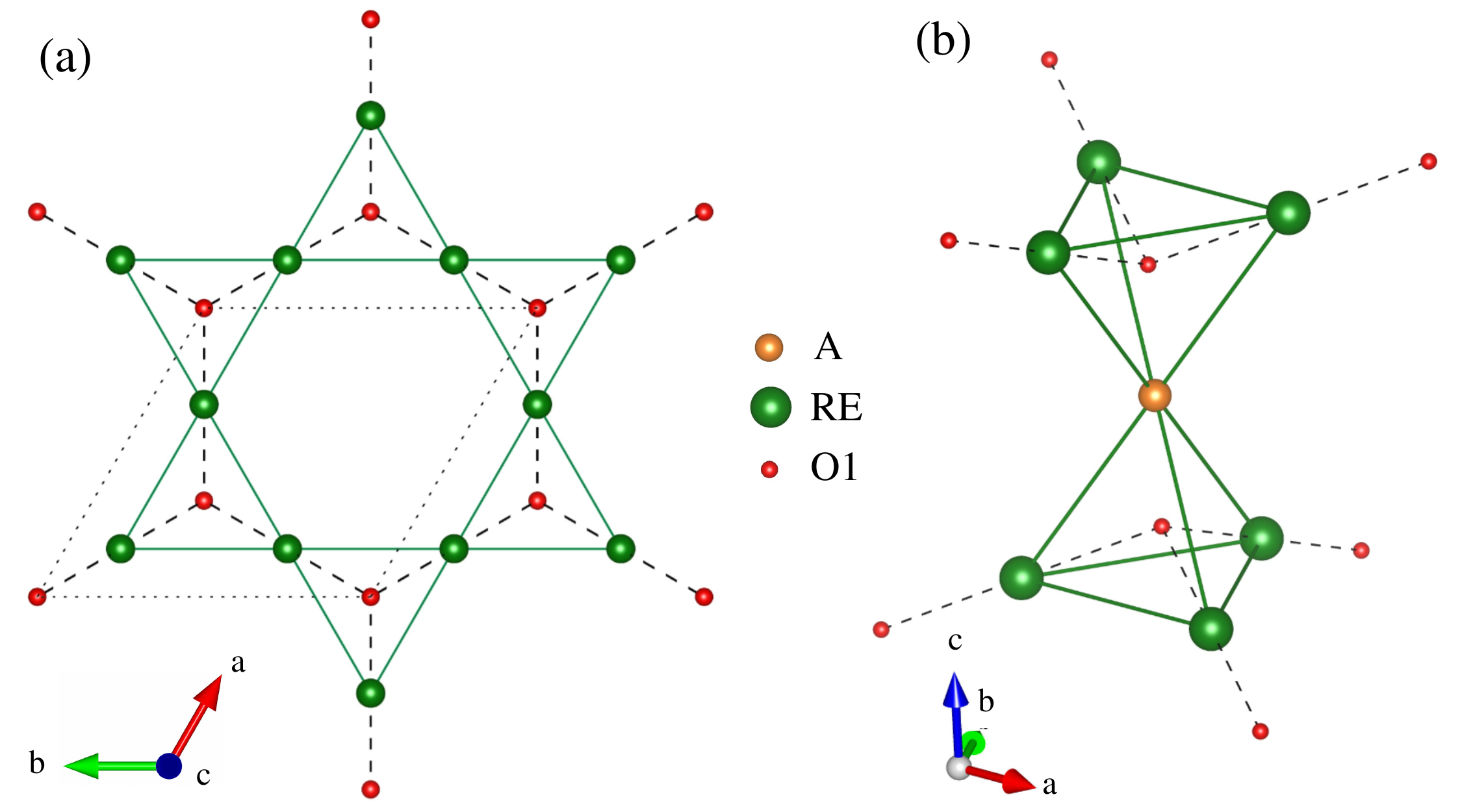}
	\end{center}
	\par
	\caption{\label{Fig:Structure}(color online) (a) A single kagome layer with surrounding O1 in a TKL A$_{2}$RE$_{3}$Sb$_{3}$O$_{14}$. (b) Illustration of substitution of A$^{2+}$ ion in a corner-shared tetrahedron for a TKL. Dashed lines represents the Tripod like local Ising axes.}
\end{figure}

\section{Structure}
As described in our previous work \cite{TKL}, the TKL is a kagome lattice based on partial ion substitution in the pyrochlore lattice. Compared with the pyrochlore lattice, one fourth of the rare earth ions are substituted by non-magnetic Mg$^{2+}$ or Zn$^{2+}$ (Fig 1.(a)), resulting in ordered kagome layers (Fig 1.(b)) that are well separated by non-magnetic triangular Mg or Zn layers and with alternating ABC stacking arrangement. More importantly, because of the similar local oxygen environment of the RE ion compared to that of the parent pyrochlore structure, we expect a similar crystal field splitting of the 4f ground state, resulting in  either Ising spins or $XY$-spin vectors that are neither uniaxial nor uniplanar. As shown in Fig. 1, there are three Ising axes for each kagome layer that are joining each RE ion to the O1 ion that are located at the center of the tetrahedron (thus the local $XY$ plane is the one that is perpendicular to the local Ising axes). It is this lack of a unique crystal axis that is neither parallel or perpendicular to the individual Ising axes of the distorted RE polyhedra that warrants use of the modifier ``Tripod" to avoid confusion with uniaxial (coplanar) kagome lattices. Such tripod-like axes will also be crucial for understanding of the low temperature magnetism for each RE-TKL. 
     
\begin{figure}[tbp]
	\linespread{1}
	\par
	\includegraphics[width= \columnwidth]{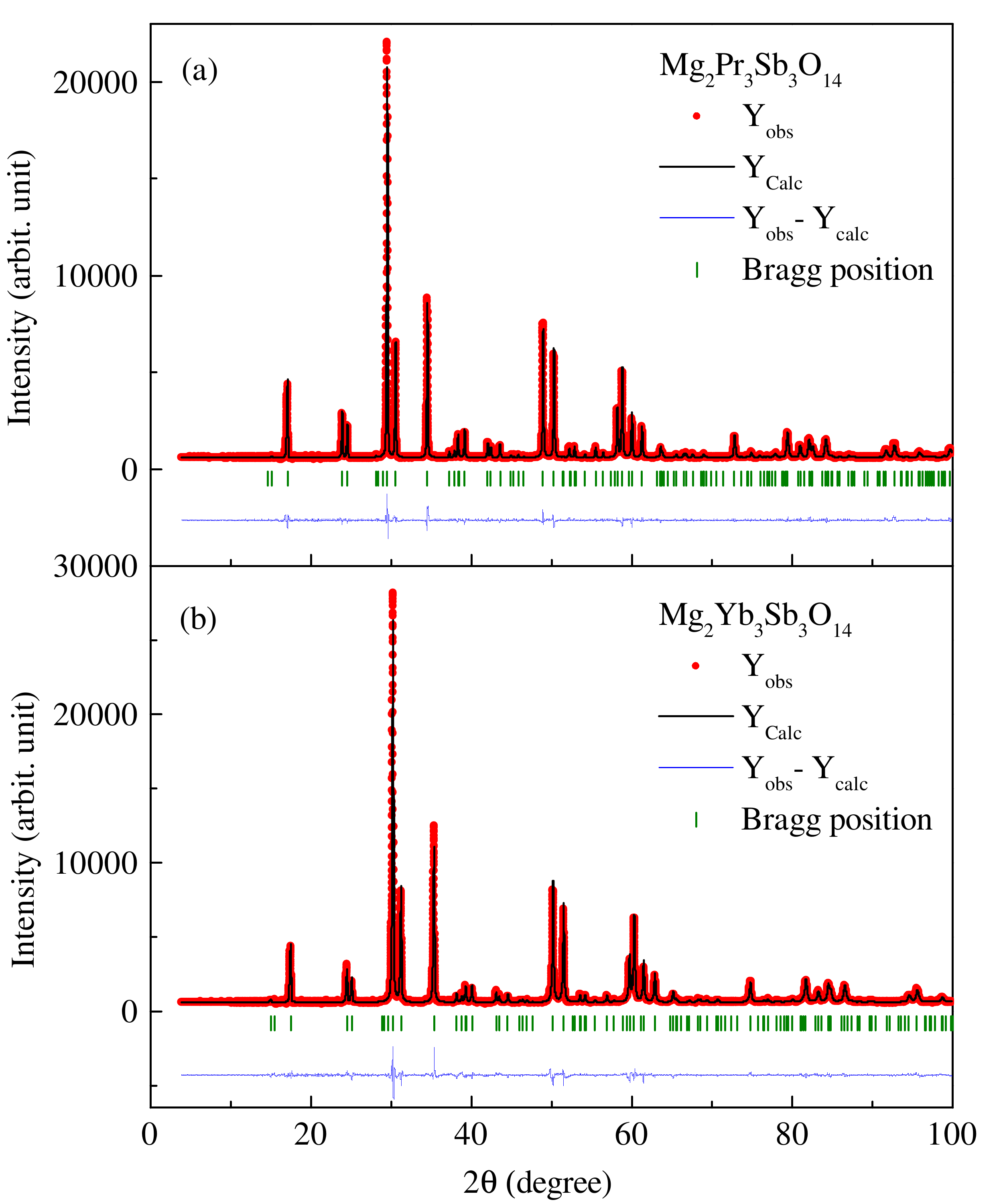}
	\par
	\caption{\label{Fig:Mg_XRD}(color online) (a) XRD patterns and best fits from Rietveld refinement for (a) \textit{MgPr} and (b) \textit{MgYb}.}
\end{figure}

The TKL compounds crystallize in a rhombohedral structure with R -3m space group in hexagonal expression. A detailed crystallographic description of the TKL structure can be found elsewhere \cite{JSSC2014Co, JSSC2014Mn, TKL, CavaZn, CavaMg}.  For the A = Mg branch, XRD patterns of all eight compounds (RE = Pr, Nd, Gd, Tb, Dy, Ho, Er, Yb) can be well-fitted by the TKL structure described earlier \cite{JSSC2014Co, JSSC2014Mn, TKL, CavaZn, CavaMg}. The associated crystallographic table with selected bond lengths are listed in Table. \ref{Tab:1}. The XRD patterns for two end members, \textit{MgPr} with the largest RE ionic radius and \textit{MgYb} with smallest RE ionic radius among those we prepared, are shown in Fig. \ref{Fig:Mg_XRD}. As shown in Fig. \ref{Fig:lattice}, the lattice parameters decrease smoothly as the RE ionic radius decreases in agreement with a previous report \cite{CavaMg}.

\begin{figure}[tbp]
	\linespread{1}
	\par
	\begin{center}
		\includegraphics[width= \columnwidth]{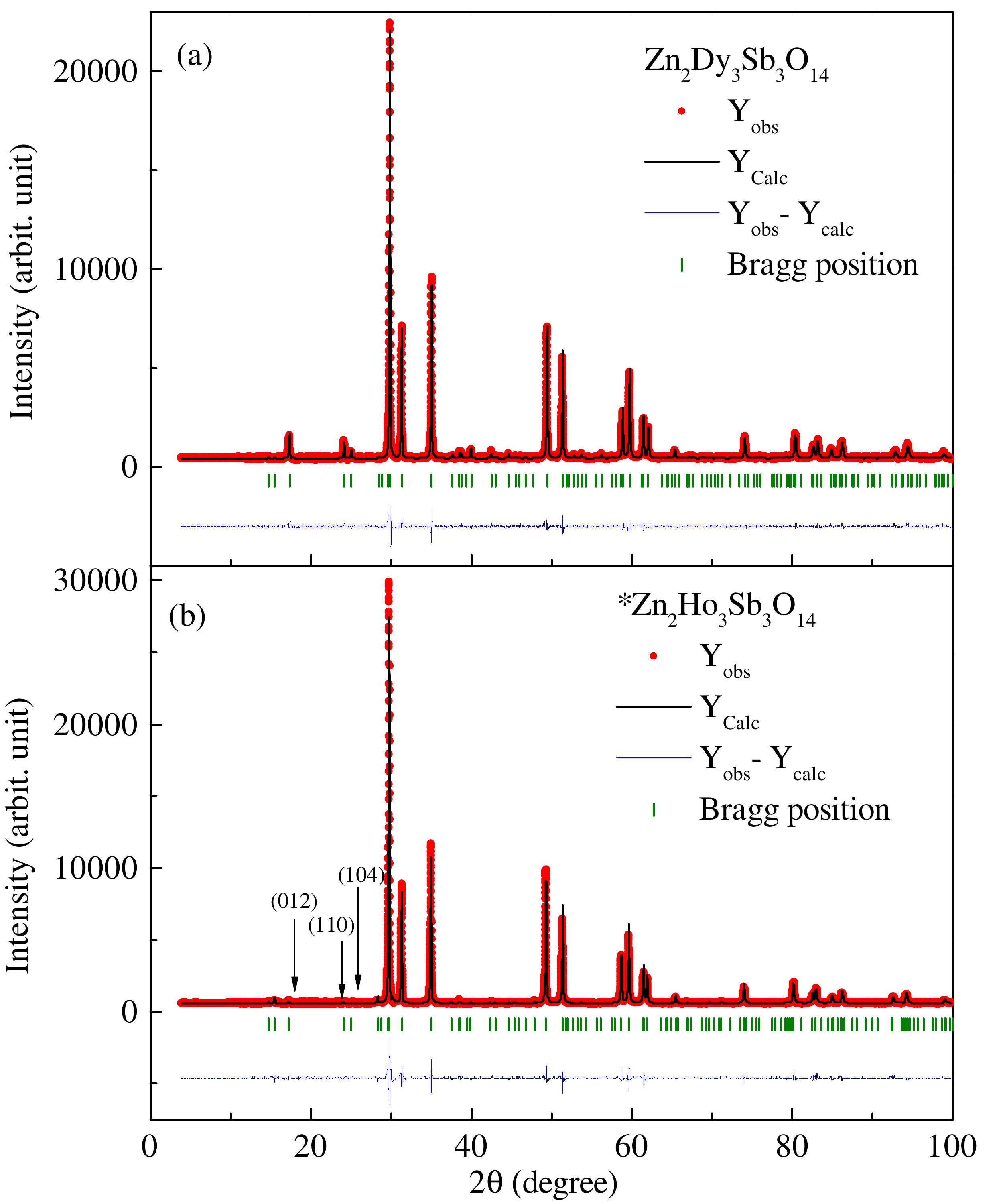}
	\end{center}
	\par
	\caption{\label{Fig:Zn_XRD}(color online) (a) XRD patterns and best fits from Rietveld refinement for two TKLs with nearby RE ions in the periodic table. (a) \textit{ZnDy} and (b)$\ast$\textit{ZnHo}. Arrows indicates where obvious discrepancies are observed.}
\end{figure}

\begin{figure}[tbp]
	\linespread{1}
	\par
	\begin{center}
		\includegraphics[width= \columnwidth ]{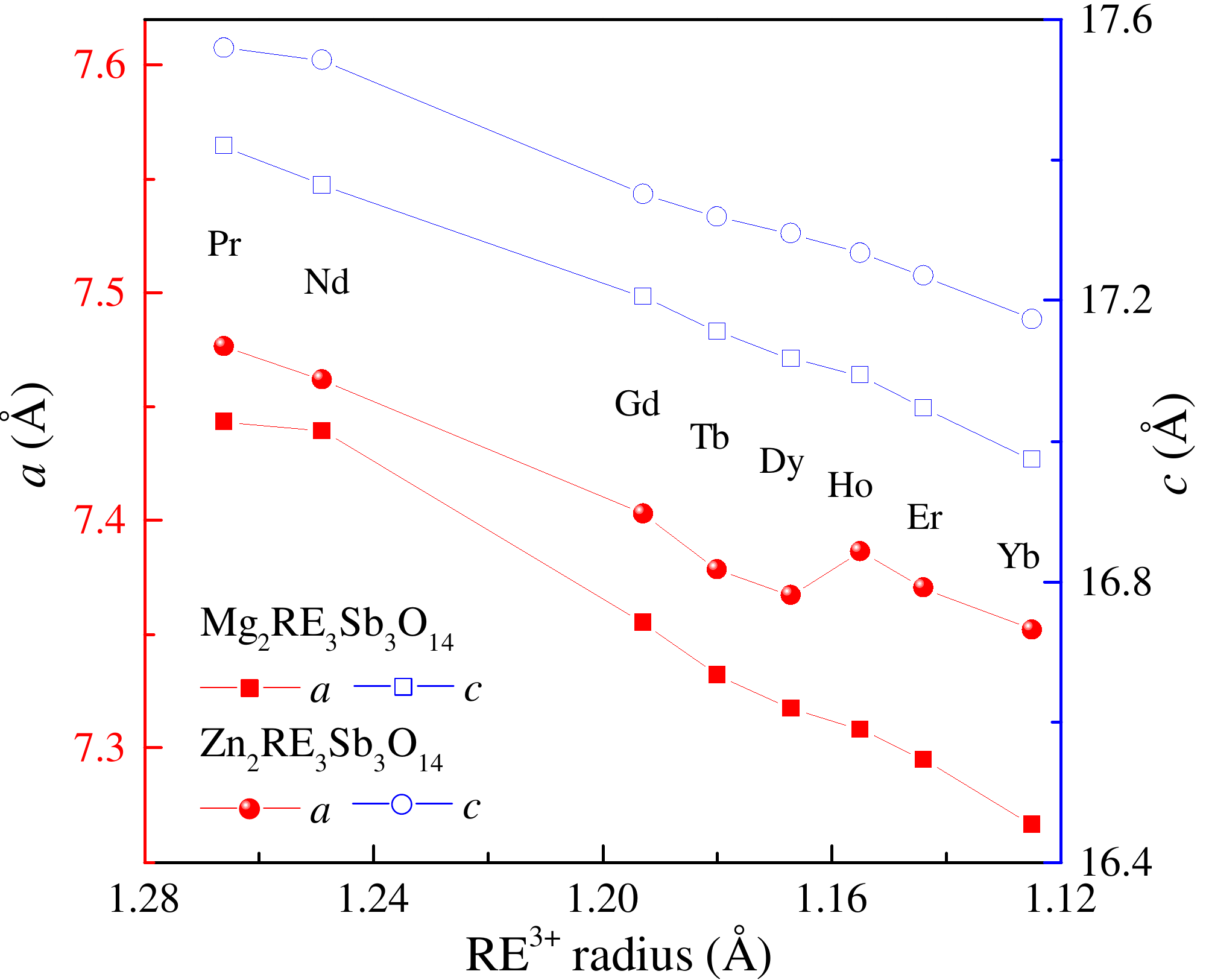}
	\end{center}
	\par
	\caption{\label{Fig:lattice}(color online) Lattice parameters obtained from Rietveld refinements as a function of RE$^{3+}$ ionic radius for all RE members in the  Mg and Zn branches of the TKL family.  }
\end{figure}

For the A = Zn branch, the XRD patterns for compounds with larger RE ionic radii (RE = Pr, Nd, Gd, Tb, Dy) closely correspond to the TKL structure while some discrepancies are observed for compounds with smaller ionic radius (RE = Ho, Er, Yb). This finding agrees with a previous report in which attempts to synthesize materials with the TKL structure based on smaller rare earth ions were unsuccessful \cite{CavaZn}. For comparison, the XRD patterns for two compounds with nearby RE$^{3+}$ ions on the periodic table, \textit{ZnDy} and \textit{ZnHo}, are plotted in Fig. \ref{Fig:Zn_XRD}. In general, the XRD pattern of \textit{ZnHo} is similar to that of \textit{ZnDy} in terms of the positions and intensity ratios of the main reflections. However, some weak reflections such as the (012), (110), and (104) Bragg peaks (marked by the arrows in Fig. \ref{Fig:Zn_XRD}(b)) observed for \textit{ZnDy} are not present in the \textit{ZnHo} data. 

The difference between \textit{ZnDy} and \textit{ZnHo} can be explained by Zn/Ho site disorder. With the similar TKL structure, it is the site mixing of Zn and Ho ions that reduces the distinctness of their original positions, increases the crystallographic symmetries, and results in a reduction of the number of Bragg reflections. Assuming total site mixing of Zn/Ho with 40\% Zn and 60\% Ho occupancy at their original Wyckoff site in a TKL, a XRD simulation will give zero intensity for the (012), (110), (104) Bragg peaks if one ignores the weak scattering from oxygen. As shown in Fig. \ref{Fig:Zn_XRD}(b), a Rietveld refinement based on a total Zn/Ho site disorder  model fits the XRD pattern reasonably well except for some small discrepancies that are likely due to unstable oxygen positions in the refinement \cite{note}. Lattice parameters for the Zn-TKL branch are shown in Fig. \ref{Fig:lattice}. A clear jump is observed between \textit{ZnDy} and \textit{ZnHo} for $a$, showing that site disorder expands the lattice within the $ab$ plane. Such site disorder is not totally unexpected when we move from lower to higher $Z$ in the 4f row. As we do so, the ionic radius of RE$^{3+}$ decreases, and finally at Ho, it becomes insufficient to be distinguished from the Zn$^{2+}$ ions when RE goes beyond during the sample synthesis at high temperature. Similar behavior has been reported for Ca$_{2}$RE$_{3}$Sb$_{3}$O$_{14}$ where the Ca/RE site disorder is present for all RE compounds of the Ca branch \cite{CaTKL}.  In order to distinguish these site-disordered lattices from other site-ordered ones, we will add a notation ``$\ast$'' before the chemical formula for disordered lattice (e.g. $\ast$\textit{ZnHo}) in the following sections. It is also possible that some slight A/RE disorder exists in the other compounds. Within the experimental resolution, the refinements based on our XRD patterns generally give 1-5\% A/RE site disorder for other TKL members of the Zn branch (RE = Pr, Nd, Gd, Tb, Dy) and all TKLs of the Mg branch.

Similar to the disorder effects found in other frustrated magnets, such A/RE site disorder in the TKL structure is likely to modify the spin-spin correlations, and tune the fragile low temperature ground state. As shown below, the total Zn/Er disorder in $\ast$\textit{ZnEr} results in a spin glass (SG) ground state.  It is also noteworthy that a small level of site disorder is likely to play an important role. For example, by comparing two separate reports of \textit{MgDy} \cite{TKL, DyECO}, the difference of the ECO ordering temperatures ($\sim$ 0.3 K in Ref. \cite{DyECO}, $\sim$ 0.37 K in Ref. \cite{TKL}) and the sharpness of the transition peaks in $C(T)$ can be attributed to the different percentages of Mg/Dy site disorder which might come from different reaction environments during sample synthesis.

\begin{table*}
	\footnotesize
	\begin{center}
		\caption{ \label{Tab:1} Summary of room temperature XRD pattern refinements for A$_2$RE$_3$Sb$_3$O$_{14}$ (A = Mg, Zn; RE = Pr, Nd, Gd, Tb, Dy, Ho, Er, Yb)}
		\renewcommand{\arraystretch}{1.32}%
		\begin{tabular}{ccccccccc}		
			\hline 
			\small	A = Mg 	& Pr & Nd  & Gd & Tb & Dy & Ho & Er & Yb \\ 
			\hline 
			IR (RE$^{3+}$) (\AA) & 1.266 & 1.249 & 1.193  & 1.180  & 1.167 & 1.155 & 1.144 & 1.125 \\
			$a$ (\AA) & 7.44347(3) & 7.43899(8) & 7.35505(6) &  7.33201(3) & 7.31781(9) & 7.30817(8) & 7.29484(9) & 7.26659(2) \\ 
			$c$ (\AA) & 17.55855(18) & 17.54255(18) & 17.35073(14) & 17.31816(35) & 17.29602(22) & 17.26724(19) & 17.23451(21) & 17.17256(27)\\ 
			Mg1(3a)  & (0, 0, 0)   & (0, 0, 0)  & (0, 0, 0) & (0, 0, 0) & (0, 0, 0) & (0, 0, 0) & (0, 0, 0) & (0, 0, 0) \\ 
			Mg2(3b)  & (0, 0, 1/2)   & (0, 0, 1/2)   & (0, 0, 1/2)  & (0, 0, 1/2)  & (0, 0, 1/2)  & (0, 0, 1/2)  & (0, 0, 1/2)  & (0, 0, 1/2)  \\ 
			RE (9d)  & ( 1/2, 0, 1/2) & ( 1/2, 0, 1/2) & ( 1/2, 0, 1/2) & ( 1/2, 0, 1/2)& ( 1/2, 0, 1/2) & ( 1/2, 0, 1/2) & ( 1/2, 0, 1/2) & ( 1/2, 0, 1/2)  \\ 
			Sb (9e)  & (1/2, 0, 0) & (1/2, 0, 0) & (1/2, 0, 0) & (1/2, 0, 0) & (1/2, 0, 0) & (1/2, 0, 0) & (1/2, 0, 0) & (1/2, 0, 0)\\
			O1 (6c)  & (0, 0, $z$) & (0, 0, $z$) &  (0, 0, $z$) &  (0, 0, $z$) &  (0, 0, $z$) &  (0, 0, $z$) &  (0, 0, $z$) &  (0, 0, $z$) \\ 
			$z$ & 0.1043(5) & 0.1078(4) & 0.1031(5) & 0.0986(6) & 0.0940(5) & 0.1085(5) & 0.1152(5) & 0.1175(5) \\
			O2 (18h) & ($x$, $\bar{x}$, $z$) & ($x$, $\bar{x}$, $z$) & ($x$, $\bar{x}$, $z$) & ($x$, $\bar{x}$, $z$) & ($x$, $\bar{x}$, $z$) & ($x$, $\bar{x}$, $z$) & ($x$, $\bar{x}$, $z$) & ($x$, $\bar{x}$, $z$)\\
			$x$ & 0.5293(4) & 0.5275(3) & 0.5320(3) & 0.5344(4) & 0.5372(3) & 0.5323(4) &   0.5357(3) & 0.5249(4)  \\
			$z$ & 0.8907(3) & 0.8914(2) & 0.8951(3) & 0.8964(3) & 0.8980(3) & 0.8968(3) & 0.8983(3) & 0.8959(3) \\
			O3 (18h) & ($x$, $\bar{x}$, $z$) & ($x$, $\bar{x}$, $z$) & ($x$, $\bar{x}$, $z$) & ($x$, $\bar{x}$, $z$) & ($x$, $\bar{x}$, $z$) & ($x$, $\bar{x}$, $z$) & ($x$, $\bar{x}$, $z$) & ($x$, $\bar{x}$, $z$)\\
			$x$ & 0.4703(4) & 0.4708(3) &  0.4754(4) & 0.4753(5) & 0.4765(4) & 0.4769(4) & 0.4751(3) & 0.4719(4)  \\
			$z$ & 0.3566(3) & 0.3558(2) &  0.3578(2) & 0.3591(3) & 0.3586(2) & 0.3596(2) & 0.3580(2) & 0.3523(2) \\
			RE-O1 (\AA) & 2.412(4) & 2.383(4) & 2.393(4) & 2.423(6) & 2.458(5)   & 2.337(4) & 2.285(4) & 2.261(4) \\
			RE-O2 (\AA) & 2.566(4) & 2.556(2) & 2.587(3) & 2.606(4) & 2.634(3)   & 2.586(4) & 2.616(3)  & 2.513(3) \\
			RE-O3 (\AA) & 2.547(6) & 2.557(4) & 2.487(4) & 2.460(6) & 2.462(4)   & 2.442(4) & 2.467(4) & 2.561(6) \\
			Intralayer RE-RE & 3.72174(4) & 3.71950(5) & 3.67753(4) & 3.66601(8) & 3.65891(5) & 3.65409(5) & 3.64742(5) & 3.63330(6) \\
			Interlayer RE-RE & 6.23482(6) & 6.22936(6) & 6.16099(5) & 6.14852(9) & 6.14016(7) & 6.13020(6) & 6.11863(7) & 6.09645(10) \\
			Overall B (${\AA}^2$) & 1.39(1) & 1.55(1) & 1.45(1) & 1.51(2) & 1.43(1) & 1.58(1) &  1.55(1)  & 1.48(2) \\ 
			R$_p$		& 3.24  & 2.05 & 2.32 & 3.65 & 2.93 & 3.37  & 3.54 & 2.94  \\
			R$_{wp}$ 	& 3.55  & 2.05 & 3.42 & 4.56 & 3.43 & 3.72  & 3.31 & 3.16 \\
			${\chi}^2$  & 2.62  & 1.15 & 1.17 & 1.51 & 1.24  & 3.89  & 3.60 & 2.27  \\ 
			\hline		
			\hline		
			\small	A = Zn 	& Pr & Nd  & Gd & Tb & Dy & Ho & Er & Yb \\
			\hline 
			IR (RE$^{3+}$) (\AA) & 1.266 & 1.249 & 1.193  & 1.180  & 1.167 & 1.155 & 1.144 & 1.125 \\
			$a$ (\AA) & 7.47622(9) & 7.46151(10) & 7.40270(11) & 7.378569(11) & 7.36714(12)  & 7.38639(3) & 7.37040(3) & 7.35212(2) \\ 
			$c$ (\AA) & 17.42042(21) & 17.36332(22)  & 17.20519(26) & 17.15565(26) & 17.11680(29) & 17.09436(7)  & 17.04657(7) & 16.97254(6) \\ 
			Zn1(3a)  & (0, 0, 0)   & (0, 0, 0)  & (0, 0, 0) & (0, 0, 0) & (0, 0, 0) & disorder & disorder & disorder \\ 
			Zn2(3b)  & (0, 0, 1/2)  & (0, 0, 1/2)  & (0, 0, 1/2)  & (0, 0, 1/2)  & (0, 0, 1/2) & disorder & disorder & disorder  \\ 
			RE (9d)  & ( 1/2, 0, 1/2) & ( 1/2, 0, 1/2) & ( 1/2, 0, 1/2) & ( 1/2, 0, 1/2)& ( 1/2, 0, 1/2) & disorder & disorder & disorder  \\ 
			Sb (9e)  & (1/2, 0, 0) & (1/2, 0, 0) & (1/2, 0, 0) & (1/2, 0, 0) & (1/2, 0, 0) & (1/2, 0, 0) & (1/2, 0, 0) & (1/2, 0, 0)\\
			O1 (6c)  & (0, 0, $z$) & (0, 0, $z$) &  (0, 0, $z$) &  (0, 0, $z$) &  (0, 0, $z$) &  (0, 0, $z$) &  (0, 0, $z$) &  (0, 0, $z$) \\ 
			$z$ & 0.1105(7) & 0.1063(8) & 0.1084(8) & 0.1061(8) & 0.1057(9) & - & - & - \\
			O2 (18h) & ($x$, $\bar{x}$, $z$) & ($x$, $\bar{x}$, $z$) & ($x$, $\bar{x}$, $z$) & ($x$, $\bar{x}$, $z$) & ($x$, $\bar{x}$, $z$) & ($x$, $\bar{x}$, $z$) & ($x$, $\bar{x}$, $z$) & ($x$, $\bar{x}$, $z$)\\
			$x$ & 0.5183(5) & 0.5206(5) & 0.5180(5) & 0.5212(5) & 0.5178(6) & - & - & - \\
			$z$ & 0.8890(3) & 0.8902(4) & 0.8888(3) & 0.8876(4) & 0.8857(4) & - & - & -  \\
			O3 (18h) & ($x$, $\bar{x}$, $z$) & ($x$, $\bar{x}$, $z$) & ($x$, $\bar{x}$, $z$) & ($x$, $\bar{x}$, $z$) & ($x$, $\bar{x}$, $z$) & ($x$, $\bar{x}$, $z$) & ($x$, $\bar{x}$, $z$) & ($x$, $\bar{x}$, $z$)\\
			$x$ & 0.4654(4) & 0.4644(4) & 0.4676(4) & 0.4662(4) & 0.4668(5) & - & - & -  \\
			$z$ & 0.3510(3) & 0.3554(3) & 0.3546(3) & 0.3566(3) & 0.3544(3) & - & - & -  \\
			RE-O1 (\AA) & 2.370(6) & 2.395(7) & 2.360(6) & 2.370(7) & 2.369(7) &- &- &-  \\
			RE-O2 (\AA) & 2.483(4) & 2.502(4) & 2.454(4) & 2.461(4) & 2.420(4) &- &- &-  \\
			RE-O3 (\AA) & 2.634(6) & 2.553(6) & 2.536(6) & 2.496(6) & 2.528(6) &- &- &-  \\
			Intralayer RE-RE & 3.73811(3) & 3.73076(3) & 3.70135(3)  & 3.68928(6) & 3.68357(7) &- &- &-  \\
			Interlayer RE-RE & 6.19490(7) & 6.17561(7) & 6.12026(10) & 6.10237(10) & 6.08907(10) &- &- &-  \\
			B (${\AA}^2$) & 1.54(1) & 1.52(1) & 1.53(1)  & 1.64(1) & 1.66(1) & 2.42(2) & 2.16(1) & 2.17(1) \\ 
			R$_p$		& 3.61  & 3.11 & 2.76  & 3.36 & 3.28  & 3.93 & 4.81 & 3.61 \\
			R$_{wp}$ 	& 5.76  & 4.92 & 5.78  & 5.48 & 5.32  & 8.68 & 9.79 & 9.73 \\
			${\chi}^2$  & 2.00  & 2.05 & 1.46  & 2.45 & 1.57  & 2.89 & 5.28 & 6.17 \\ 
			\hline		
		\end{tabular}  \quad
	\end{center}
\end{table*}

\section{Magnetic properties}

\begin{figure*}[tbp]
	\linespread{1}
	\par
	\begin{center}
		\includegraphics[width= 6.8 in ]{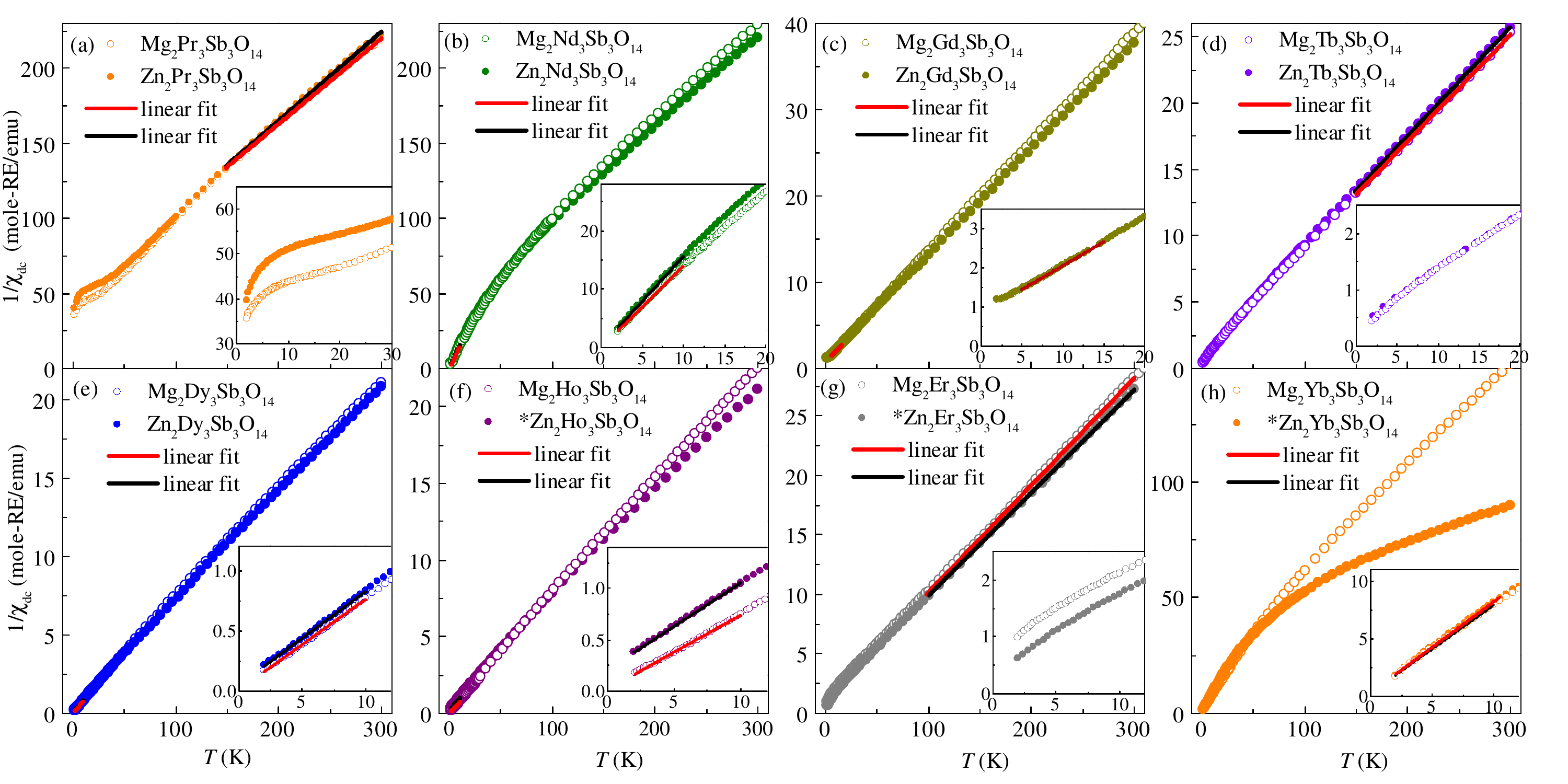}
	\end{center}
	\par
	\caption{\label{Fig:Chi}(color online) (a-h) Inverse $\chi_{dc}$ from 2 K to  300 K for all A$_{2}$RE$_{3}$Sb$_{3}$O$_{14}$ (A = Mg, Zn; RE = Pr, Nd, Gd, Tb, Dy, Ho, Er, Yb) compounds. Insets:  1/$\chi_{dc}$ at low temperature regions.}
\end{figure*}

\subsection{\textit{MgPr} and \textit{ZnPr}}
For \textit{MgPr}, a Curie-Weiss (CW) fit of 1/$\chi_{dc}$ from 150 K to 300 K (Fig. \ref{Fig:Chi}(a)),  yields a Weiss temperature, $\theta_{W}$ = -46.18 K and an effective magnetic moment, $\mu_{eff}$ = 3.40 $\mu_{B}$. For \textit{ZnPr}, a similar fit yields $\theta_{W}$ = -68.43 K and $\mu_{eff}$ = 3.61 $\mu_{B}$. These $\mu_{eff}$ values are consistent with the free ion moment of $\mu_{eff}$ = 3.58 $\mu_{B}$ expected for Pr$^{3+}$ ions. Below 50 K, 1/$\chi_{dc}$ becomes flat, followed by another slop change below 10 K (Fig. \ref{Fig:Chi}(a) inset), suggesting changes of magnetic moments and spin-spin interactions in these temperature region due to CEF effects.

\begin{figure}[tbp]
	\linespread{1}
	\par
	\begin{center}
		\includegraphics[width= \columnwidth ]{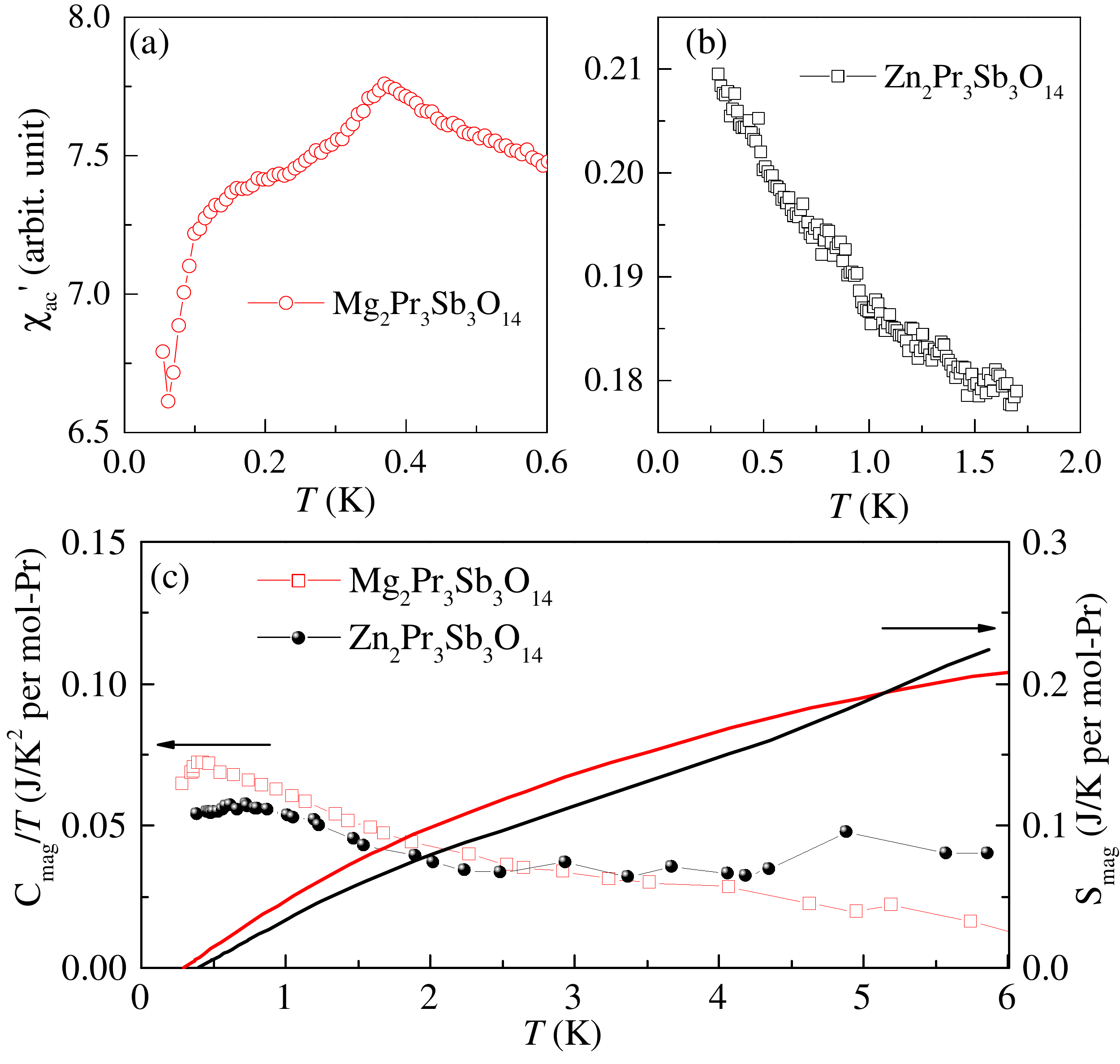}
	\end{center}
	\par
	\caption{\label{Fig:Pr}(color online) Temperature dependence of the real part of $\chi_{ac}$ for (a) \textit{MgPr}, and (b) ZnPr. (c) $C_{mag}/T$ and $S_{mag}$ for both compounds.}
\end{figure}

For \textit{MgPr}, a broad feature around 0.35 K is observed at zero field $\chi_{ac}'$ (Fig. \ref*{Fig:Pr}(a)) while no obvious sign of LRO is observed from $\chi_{ac}'$ down to 50 mK. For \textit{ZnPr} (Fig. \ref*{Fig:Pr}(b)), LRO is also not observed in $\chi_{ac}'$ where a Curie-Weiss type behavior is persistent down to the lowest measured temperature of 0.3 K. For both \textit{MgPr} and \textit{ZnPr}, the absolute values of $C_{mag}/T$ below 10 K is extremely small ($<$ 0.1 $J/K^2$ per mol-Pr), in sharp contrast to that observed in Pr-pyrochlores within the same temperature range ($\sim$ 2 $J/K^2$ per mol-Pr) \cite{PrSn, PrZr}. The integrated magnetic entropy ($S_{mag}$) from 0.35 K to 6 K recovers $\sim$ 0.2 $J/K$ per mol-Pr, a value that is 3\% of Rln2.

In the Pr-pyrochlore compound, the single-ion ground state is a non-Kramers doublet with a large Ising-like anisotropy, whose first excited crystal electric field (CEF) level is a non-magnetic singlet that is well separated from the ground state doublet (18 meV in  Pr$_2$Sn$_2$O$_7$ \cite{PrSn, PrSnCEF}, and 9.5 meV in Pr$_2$Zr$_2$O$_7$ \cite{PrZr}). Such effective spin-1/2 Ising systems with antiferromagnetic (AFM) exchange interactions give rises to a quantum spin ice ground state at low temperature \cite{PrSn, PrZr}, where spin fluctuations from a quantum superposition of the spin ice manifold suppresses LRO.

The small values of $C_{mag}$ and $S_{mag}$ observed in the Pr-TKLs suggests that the lower crystal field symmetry in the TKLs lifts the degeneracy of the low energy states probed in the pyrochlores. As discussed below, with non-Kramers Pr$^{3+}$, the local site symmetry could alter the CEF spectrum to mix the doublets and result in a non-magnetic singlet ground state.  Such a non-magnetic state is consistent with the small values of $C_{mag}$ and $S_{mag}$, as well as the flat $\chi_{dc}$ (steep $1/\chi_{dc}$) observed at low temperature (Fig. \ref{Fig:Chi} (a)).
Therefore, the signals observed in $\chi_{ac}'$ of  \textit{MgPr} and \textit{ZnPr} are likely due to a combination of Van Vleck susceptibility and a contribution from magnetic impurities that is not observed by the XRD.

\subsection{\textit{MgNd} and \textit{ZnNd}}
For \textit{MgNd}, a CW fit from 150 K to 300 K of  $1/\chi_{dc}$  yields $\theta_{W}$ = -66.36 K and $\mu_{eff}$ = 3.58 $\mu_{B}$. For \textit{ZnNd}, similar fit yields $\theta_{W}$ = -60.47 K and $\mu_{eff}$ = 3.60 $\mu_{B}$. The  $\mu_{eff}$  values are consistent with the free ion moment of $\mu_{eff}$ = 3.62 $\mu_{B}$ expected for Nd$^{3+}$. A CW fit in low temperature region (2 K-10 K, Fig. \ref{Fig:Chi}(b)) yields $\theta_{W}$ = -0.01 K, $\mu_{eff}$ = 2.38 $\mu_{B}$ for \textit{MgNd} and $\theta_{W}$ = -0.11 K, $\mu_{eff}$ = 2.28 $\mu_{B}$ for \textit{ZnNd}. These numbers are similar to that of the Nd pyrochlore \cite{NdPd,NdSn} and consistent with a recent report of the \textit{MgNd} \cite{NdTKL}.
   
\begin{figure}[tbp]
	\linespread{1}
	\par
	\begin{center}
		\includegraphics[width= \columnwidth ]{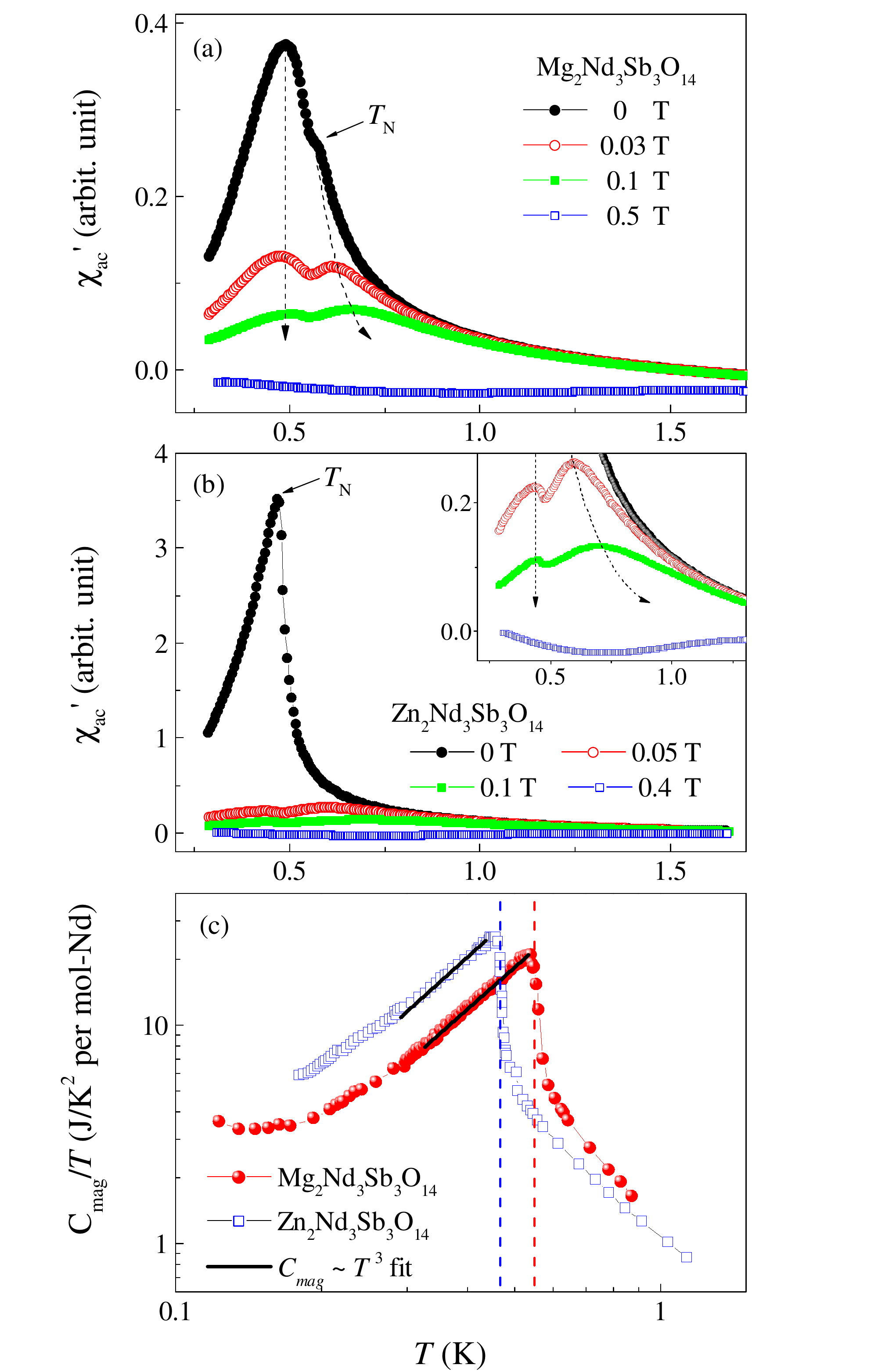}
	\end{center}
	\par
	\caption{\label{Fig:Nd}(color online) Temperature dependence of the real part of $\chi_{ac}$ under different dc fields for (a) \textit{MgNd} and (b) \textit{ZnNd}. (c) $C_{mag}/T$ on a log-log scale for \textit{MgNd} and \textit{ZnNd}. Two $C_{mag}$ $\sim$ $T^3$ fits are shown as black solid lines. }
\end{figure}

For \textit{MgNd}, a broad peak is observed in the zero field $\chi_{ac}'$ at 0.49 K, accompanied by a shoulder around 0.55 K (Fig. \ref{Fig:Nd}(a)). A small dc-field of 0.03 T reduces its height and separates the positions of these features.
With increasing fields, the position of the low temperature peak does not show obvious field dependence while the shoulder-related peak moves to higher temperatures. In Fig. \ref{Fig:Nd} (c), $C_{mag}/T$ of \textit{MgNd} shows a $\lambda$-like peak at 0.55 K, indicating a second order AFM LRO transition.
For \textit{ZnPr}, a similar peak in $C_{mag}/T$ is observed at a slightly lower temperature of 0.47 K. Accordingly, a sharp peak is observed at 0.47 K in $\chi_{ac}'$ (Fig. \ref{Fig:Nd}(a)) at zero field. Similar to that of \textit{MgNd}, this peak splits into two when a small dc field is applied (shown in Fig. \ref{Fig:Nd}(b) inset). For both compounds, the $C_{mag}$ show a $T^3$ behavior below $T_N$, consistent with linear dispersive spin wave excitations in three dimensions. 

The magnetic ground state for \textit{MgNd} has been recently studied by Scheie et. al \cite{NdTKL}, where a second order phase transition with a non-coplanar $k$ = 0 spin ordering has been revealed by elastic neutron scattering and $C(T)$ measurements. The proposed spin structure is an all-in-all-out spin state mentioned above where three spins in each triangle are pointing in or out of the local tripod directions (along RE-O1) simultaneously. Such a state also resembles the all-in-all-out spin structure for Ising spins on pyrochlore lattices \cite{AIAU}, which has been observed in Nd$_2$Sn$_2$O$_7$ \cite{NdSn}. Regarding the double peak feature observed in $\chi_{ac}'$, it is the position of the shoulder at zero field (0.55 K) that agrees with the LRO transition in $C_{mag}$. Then the appearance of the 0.47 K peak in $\chi_{ac}'$ seems to suggest a two-step transition. Additionally, a closer look of the order parameter scan of the (101) magnetic Bragg peaks from Ref. \cite{NdTKL} seems to reveal a slope change around 0.45 K. Another possibility is that, since the lower temperature peak does not appear in  $C_{mag}/T$, the two-step feature could be due to the grain effect, wherein grains with different crystal axes respond differently with respect to the applied magnetic field. For \textit{ZnNd}, two-step feature more closely converges in temperature at zero field, which
can only be distinguished in $\chi_{ac}'$ under a smaller dc field. For both compounds, further measurements under a small magnetic field will be helpful to clarify the nature of the two-step transition.

\subsection{\textit{MgGd} and \textit{ZnGd}}
The magnetic properties of \textit{MgGd} have been reported in our previous study \cite{TKL}. For \textit{ZnGd}, data from 5 K to 15 K of  $1/\chi_{dc}$ (Fig. \ref{Fig:Chi}(c)) yields $\theta_{W}$ = -6.85 K and $\mu_{eff}$ = 8.09 $\mu_{B}$, which is consistent with $\mu_{eff}$ = 7.94 $\mu_{B}$ expected for Gd$^{3+}$ ion. This $\theta_{W}$ value is slightly smaller than that of \textit{MgGd} with the same fitting range ($\theta_{W}$ = -6.70 K). 

\begin{figure}[tbp]
	\linespread{1}
	\par
	\begin{center}
		\includegraphics[width= \columnwidth ]{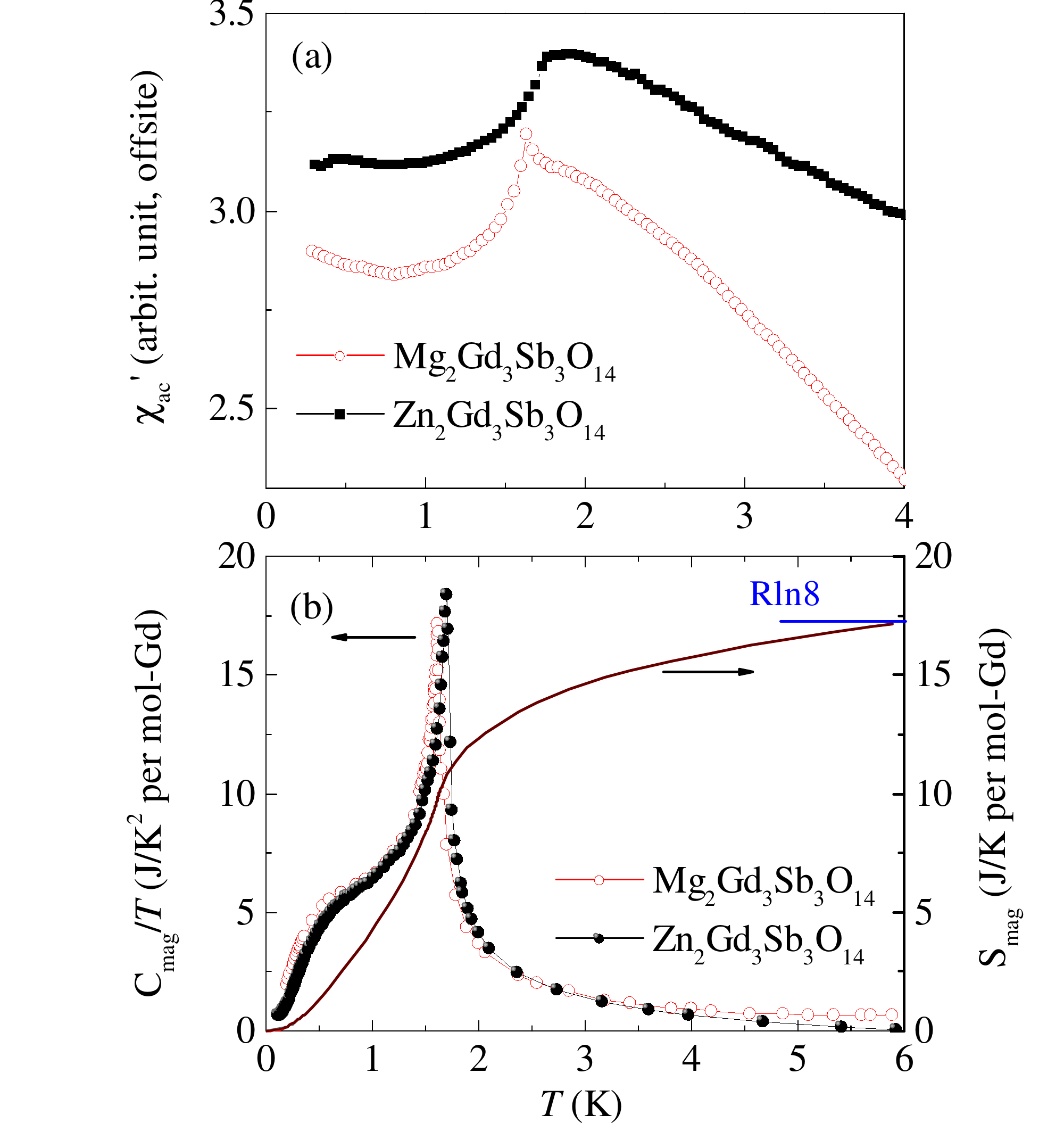}
	\end{center}
	\par
	\caption{\label{Fig:Gd}(color online) Temperature dependence of the (a) real part of $\chi_{ac}$ and (b) $C_{mag}/T$ and $S_{mag}$ for \textit{MgGd} and \textit{ZnGd}.}
\end{figure}

For \textit{ZnGd}, $\chi_{ac}'$ shows an inflection point around 1.69 K, which is consistent with a sharp peak at the same temperature observed on $C_{mag}/T$ (Fig. \ref{Fig:Gd}). The integrated $S_{mag}$ from 0.2 K to 6 K reaches 17.16 J/K per mole-Gd, consistent with Rln8 = 17.29 J/K expected for a Gd-ion.  All these features are similar to those of \textit{MgGd} \cite{TKL}, indicating that LRO occurs in both compounds. As predicted by the Luttinger-Tisza theory in the previous study \cite{TKL, Dipoles}, the ground state should be the 120 $^{\circ}$ spin structure, due to the strong dipole-dipole interaction. The ordering in \textit{ZnGd} once again demonstrates the importance of dipolar interactions for the 2D Heisenberg system.

\subsection{\textit{MgTb} and \textit{ZnTb}}
For \textit{MgTb}, a CW fit from 150 K to 300 K (Fig. \ref{Fig:Chi}(d)) of  $1/\chi_{dc}$ yields a $\theta_{W}$ = -13.70 K and $\mu_{eff}$ = 9.98 $\mu_{B}$. The $\theta_{W}$ value is slightly larger than previously reported \cite{CavaMg}. For \textit{ZnTb}, a similar fit yields  $\theta_{W}$ = -13.41 K and $\mu_{eff}$ = 9.86 $\mu_{B}$. The effective moments are slightly larger than the free ion moment of $\mu_{eff}$ = 9.72 $\mu_{B}$ expected for Tb$^{3+}$.

\begin{figure}[tbp]
	\linespread{1}
	\par
	\begin{center}
		\includegraphics[width= \columnwidth  ]{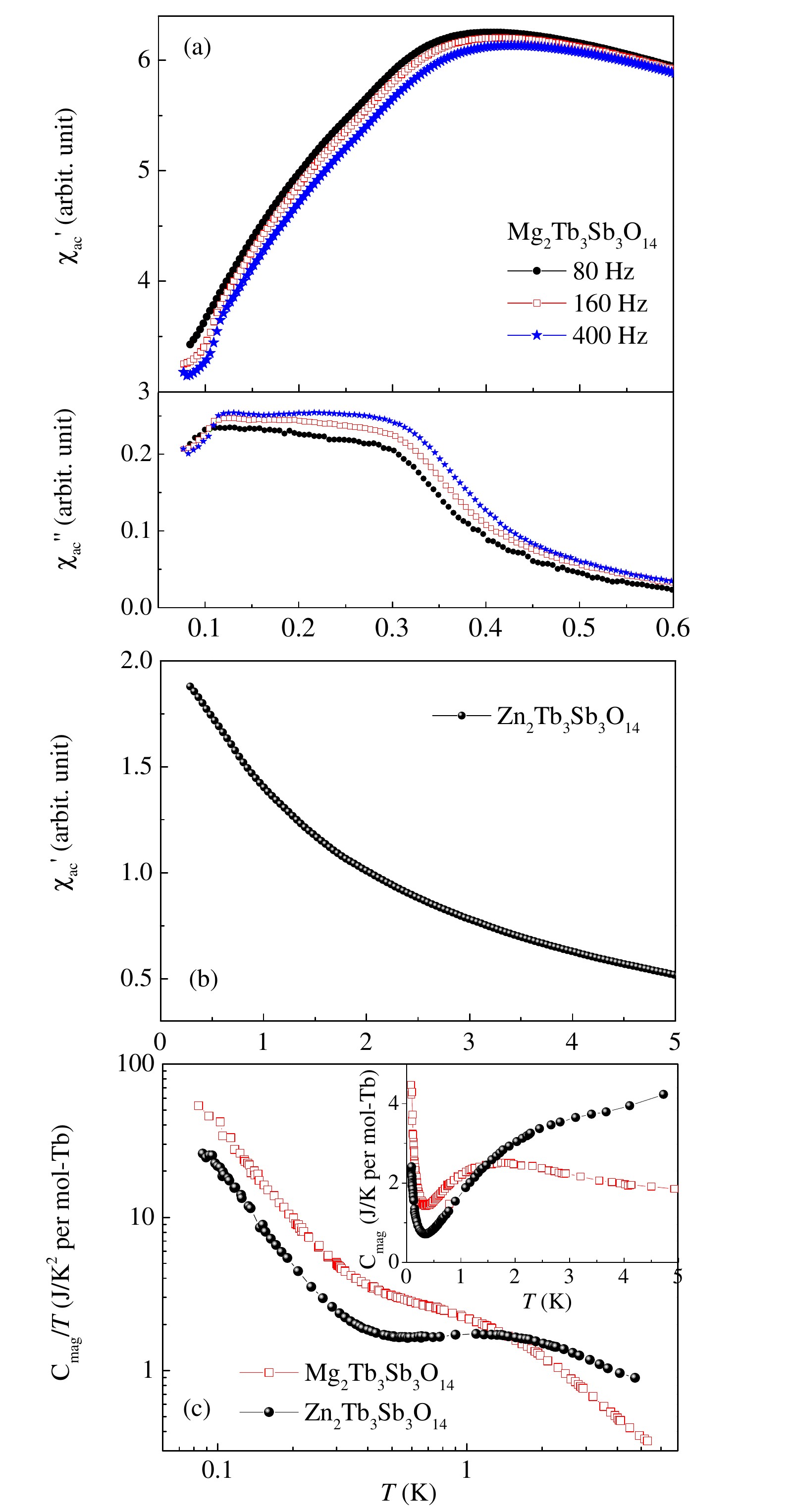}
	\end{center}
	\par
	\caption{\label{Fig:Tb}(color online) (a) Temperature dependence of  the real part of $\chi_{ac}$ with different ac field frequencies for \textit{MgTb}. (b) The real part of $\chi_{ac}$ for \textit{ZnTb}. (c) Low temperature $C_{mag}/T$ in log-log scale for \textit{MgTb} and \textit{ZnTb}. Inset: $C_{mag}$ plots in linear scale.}
\end{figure}

For \textit{MgTb}, both the real ($\chi_{ac}'$) and imaginary  ($\chi_{ac}''$) part of ac susceptibility show a broad feature around 0.35 K that is frequency dependent. Below this temperature, an anomaly is observed at 0.12 K that is more clearly seen in $\chi_{ac}''$. Aside from these, no sharp LRO feature is observed down to 50 mK (Fig. \ref{Fig:Tb}(a)). For \textit{ZnTb}, $\chi_{ac}'$ indicates a paramagnetic behavior down to the lowest measured temperature of 0.3 K, which is frequency independent (Fig. \ref{Fig:Tb}(b)). For both compounds, the absence of LRO is further confirmed by the specific heat measurement where no singularity is observed down to 80 mK. Instead, the $C_{mag}$ shows a broad feature (around 1.5 K for the \textit{MgTb} and 2.5 K for \textit{ZnTb}), followed by a power law rise below 0.3 K, as seen in the log-log plot of Fig. 9(c). We attribute this behavior to the nuclear spin degree of freedom of Tb$^{3+}$ ion ($^{159}$Tb, nuclear spin $I$ = 3/2).

The absence of LRO in  \textit{MgTb} and  \textit{ZnTb}  is reminiscent of similar behavior in the pyrochlore compound Tb$_2$Ti$_2$O$_7$.  There, the Tb$^{3+}$ ions have Ising spins coupled through AFM interactions with $\theta_{W}$ = -19 K. Yet no LRO is observed down to 0.1 K \cite{TbTi}, making it a good candidate for hosting the  QSL state. Later works indicate a Coulomb phase and short-range ordering (SRO) spin ice correlations \cite{TbTi2,TbTi3,TbTi4} with strong spin lattice coupling at low temperatures \cite{TbTi5}. Theory has invoked virtual transitions between CEF levels of the ground state and the excited state doublets, which precludes conventional order \cite{TbTi6,TbTi7} but can lead to a QSL state. Due to the similar local environment, we expect the same Ising anisotropy and similar low lying CEFs in Tb-TKLs as well. It is possible that a similar virtual transition is playing an important role to obstruct LRO in Tb-TKLs, making them promising QSL candidates.

\subsection{\textit{MgDy} and \textit{ZnDy}}
The magnetic properties for \textit{MgDy} have been reported in our previous study \cite{TKL}. For \textit{ZnDy}, a Curie-Weiss fit from 2 K to 10 K (Fig. \ref{Fig:Chi} (e)) of  $1/\chi_{dc}$ yields $\theta_{W}$ = -0.72 K and $\mu_{eff}$ = 10.20 $\mu_{B}$. Both values are similar to those of \textit{MgDy} ($\theta_{W}$ = -0.18 K and $\mu_{eff}$ = 10.20 $\mu_{B}$). The effective moments are consistent with the large free ion moment of $\mu_{eff}$ = 10.63 $\mu_{B}$ expected for Dy$^{3+}$. The small negative value of $\theta_{W}$ suggests the competition between the ferromagnetic dipolar interactions and AFM exchange couplings at low temperature. 

\begin{figure}[tbp]
	\linespread{1}
	\par
	\begin{center}
		\includegraphics[width= \columnwidth ]{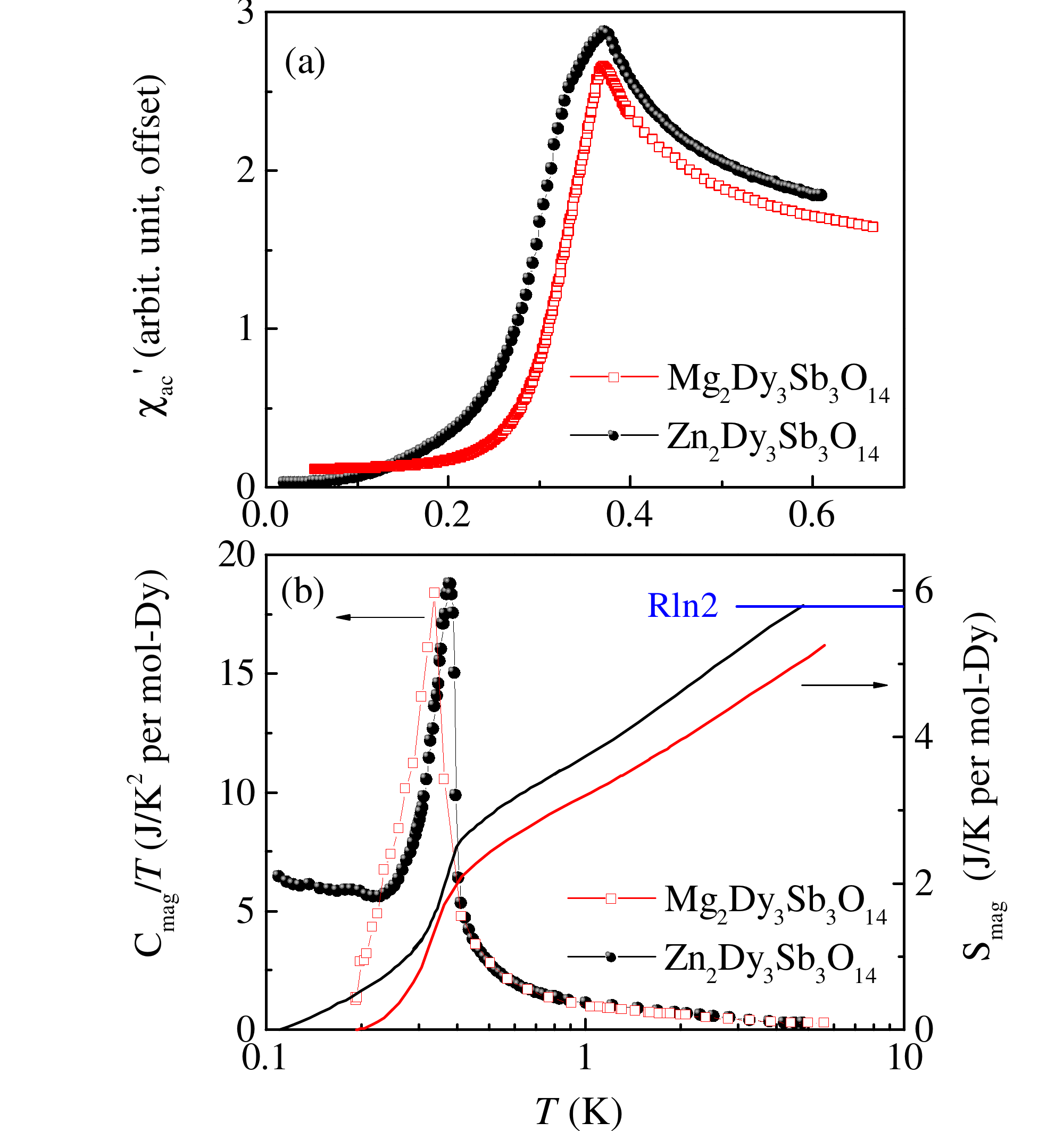}
	\end{center}
	\par
	\caption{\label{Fig:Dy}(color online) (a) Temperature dependence of the real part of $\chi_{ac}$ for \textit{MgDy} and \textit{ZnDy}. (b) Low temperature $C_{mag}/T$ and $S_{mag}$ for the two Dy-TKLs.}
\end{figure}

As shown in Fig. \ref{Fig:Dy}, similar to that of \textit{MgDy}, sharp peaks with an inflection point at  0.39 K are observed in both $\chi_{ac}'$ (which is frequency independent) and  $C_{mag}/T$ for \textit{ZnDy}. An extra increase is observed in $C_{mag}/T$ of \textit{ZnDy} below 0.2 K that is absent in that of \textit{MgDy}. Due to this increase, the integrated $S_{mag}$ from 0.1 K to 6 K reaches 5.81 J/K per mole Dy, which is slight larger than that of \textit{MgDy} ($S_{mag}$ = 5.38 J/K per mole Dy \cite{TKL}).  Besides this, the position, intensity, and shape of the $C_{mag}$ peak are almost identical for \textit{MgDy} and \textit{ZnDy}. 

For \textit{MgDy}, the transition at 0.37 K was first understood as a LRO of the Dy$^{3+}$ spins because of the lack of frequency dependence of $\chi_{ac}'$ and the sharpness of the transition in both $\chi_{ac}'$, and $C_{mag}$. Recent neutron scattering experiments \cite{DyECO} illustrate that this LRO transition is actually an ECO where emergent magnetic charge degrees of freedom exhibit LRO while spins remain partially disordered. Thus, the partially disordered spins give an averaged LRO moment on each Dy site, which is the origin of the LRO features observed in  $\chi_{ac}'$ and $C_{mag}$.  With almost identical behaviors observed here between \textit{MgDy} and \textit{ZnDy}, it is likely that \textit{ZnDy} shares the same ECO ground state as well. In this sense, the increase of $C_{mag}$ below 0.25 K in \textit{ZnDy} is unexpected. This signal is not likely due to the Dy nuclear spin given the same magnetic ion and same hyperfine coupling. One possibility is that it is related to the additional spin dynamics below the ECO, which has been proposed theoretically by Monte Carlo simulations \cite{ECO1,DyECO}. Since the ECO state is a spin partially-ordered state, it contains a non-zero entropy density of 0.108R = 0.90 J/K per mole spin.  A fully spin-ordered state with zero entropy density could be achieved at an even lower temperature given a non-local (ring flip) spin dynamics that does not cost energies for an ECO state. The extra entropy recovered in \textit{ZnDy} compared to that of \textit{MgDy} ($\sim$ 0.4 J/K per mole-Dy) is consistent with such a picture where spin dynamics drives the system further towards LRO and fully recovers the total entropy of Rln2. Yet it remains unknown why such dynamics is absent in \textit{MgDy}.

\subsection{\textit{MgHo} and $\ast$\textit{ZnHo}}
For \textit{MgHo}, a low temperature CW fit from 2 K to 10 K of  $1/\chi_{dc}$ (Fig. \ref{Fig:Chi} (f)) yields  $\theta_{W}$ = -0.27 K and $\mu_{eff}$ = 10.54 $\mu_{B}$.  For $\ast$\textit{ZnHo}, a similar fit (Fig. \ref{Fig:Chi} (f)) yields  $\theta_{W}$ = -2.49 K and a  $\mu_{eff}$ = 10.22 $\mu_{B}$. These values for the effective moment are close to the free ion moments of $\mu_{eff}$ = 10.63 $\mu_{B}$ expected for Ho$^{3+}$. The origin for the small negative $\theta_{W}$  here is the same as those of the Dy-TKLs.

\begin{figure}[tbp]
	\linespread{1}
	\par
	\begin{center}
		\includegraphics[width= \columnwidth ]{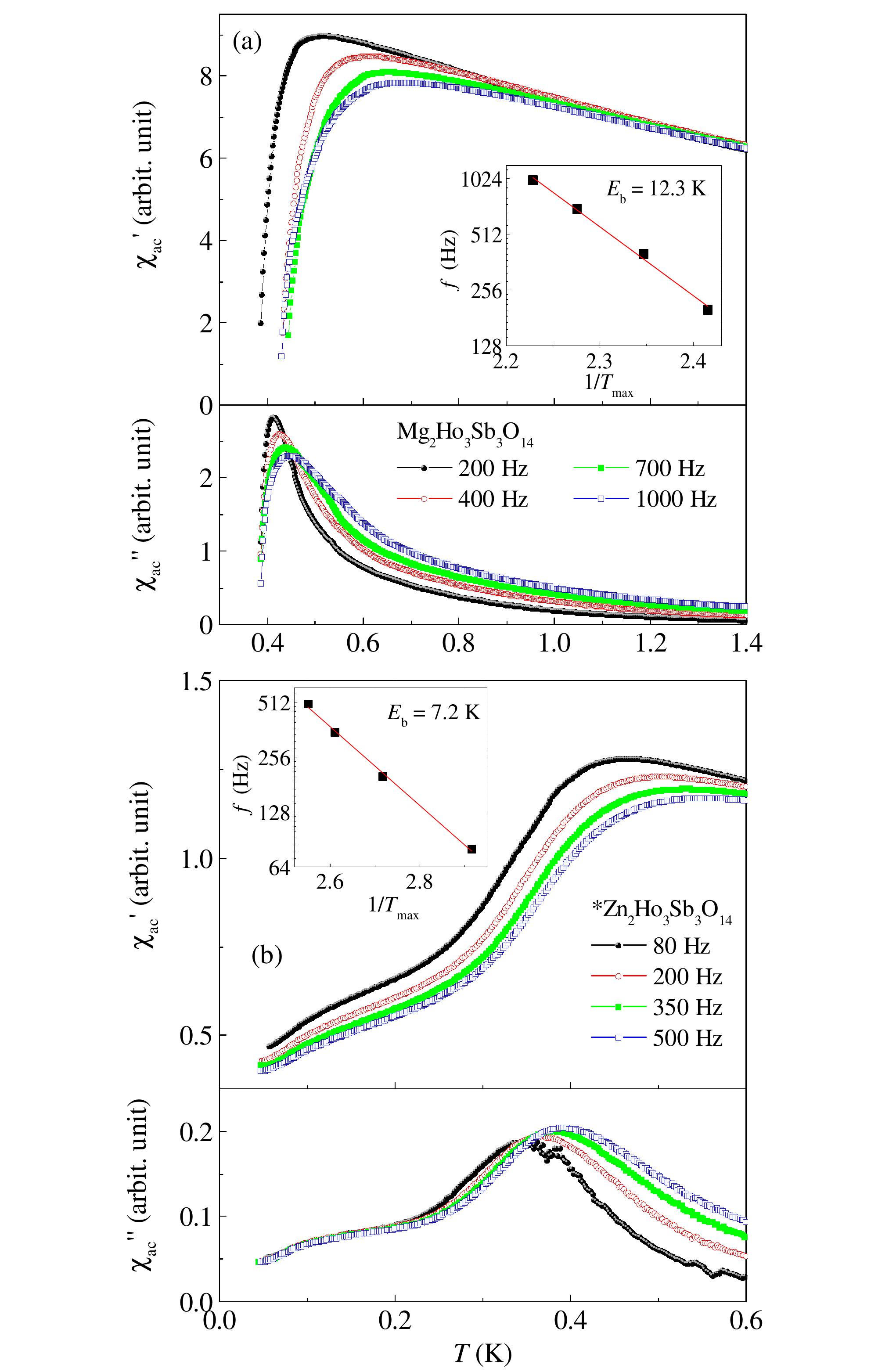}
	\end{center}
	\par
	\caption{\label{Fig:Ho}(color online) Temperature dependence of the real part ($\chi_{ac}'$ ) and imaginary part ($\chi_{ac}$'')  of $\chi_{ac}$ with different ac field frequencies for (a) \textit{MgHo} and (b) $\ast$\textit{ZnHo}. Insert: the ac field frequencies versus inverse peak maximum temperature of $\chi_{ac}''$.}
\end{figure}

For \textit{MgHo}, both the real part ($\chi_{ac}'$) and imaginary part ($\chi_{ac}''$) of $\chi_{ac}$ show a broad peak around 0.4 K with strong frequency dependence (Fig. \ref{Fig:Ho} (a)). With increasing frequency of the ac field, the peak becomes even broader and shifts higher in temperature. By fitting the ac field frequency ($f$) and $\chi_{ac}''$ peak maximum $T_{max}$ to an Arrhenius formula $f = f_0 exp(-E_b/T )$, we obtain an energy barrier of $E_b$ = 12.3 K (Fig. \ref{Fig:Ho} (a) inset). For $\ast$\textit{ZnHo}, an even broader peak with similar frequency dependent is observed in both $\chi_{ac}'$ and $\chi_{ac}''$ around 0.45 K (Fig.\ref{Fig:Ho} (a)). The corresponding energy barrier from the Arrhenius fit is $E_b$ = 7.2 K.

Similar frequency dependence of $\chi_{ac}$ have been observed in their parent spin ice pyrochlore Dy$_2$Ti$_2$O$_7$ and  Ho$_2$Ti$_2$O$_7$ above the spin freezing transition \cite{Nature1999,Science2001}. In Dy$_2$Ti$_2$O$_7$, at a temperature region above the spin freezing where the monopole density is high but where the double-monopoles are few, the relaxation behavior in $\chi_{ac}$ can be well parametrized by an Arrhenius law. The related value of  $E_b$ is equal to twice the effective spin-spin coupling ($J_{eff}$), which is actually the energy cost of a single monopole defect \cite{DTOac}. In Ho$_2$Ti$_2$O$_7$, a larger value of  $E_b$ = 13.08 K ($\sim$ 6$J_{eff}$) is observed, whose origin is not well understood \cite{Hoac}. 

Then what is the ground state and what is the associated energy barrier in the Ho-TKL? Here, we propose two possibilities. First, with close similarities of the spin anisotropies and spin-spin interactions between \textit{MgDy} and \textit{MgHo},  the similar ECO state would be expected. Indeed, for \textit{MgHo}, the 0.4 K transition in $\chi_{ac}$ is close in temperature to the 0.37 K ECO transition in \textit{MgDy}. However, the frequency dependence of the $\chi_{ac}$ clearly differentiates it from that of \textit{MgDy}, suggesting an ECO state with  extra spin relaxation process due to thermal or quantum fluctuations. As discussed below in section V, such behavior is likely related to non-Kramers nature Ho$^{3+}$, where the extra lowering of site symmetry in the TKL system splits the energy of CEF ground state doublets in Ho$^{3+}$ at a finite energy, building an energy barrier for spin-spin interactions. The related relaxation process could also involve a hyperfine contribution that is not uncommon in Ho magnets at these temperatures. 
Second, given the large dipolar couplings which act ferromagnetic exchange interactions at the nearest neighbor, \textit{MgHo} is a potential candidate for hosting the kagome spin ice  (KSI) state. As discussed in Ref. \cite{KSI} and \cite{TKL}, classical spins with TKL-like Ising anisotropy on a kagome lattice are highly frustrated which will result in a large ground state degeneracy and the zero-point entropy.   Similar to that of the pyrochlore spin ice, if the KSI state is achieved, the broad peak observed in $\chi_{ac}$ of \textit{MgHo} could represent a spin freezing process with $T_f$ $\le$ 0.4 K while the system enters the SRO state characterized by the KSI ice rule. Then the value of  $E_b$ = 12.3 K is likely related to the energy difference from the ice rule state to the excited all-in-all-out state for a single Ho-triangle. The KSI state distinguishes itself from the ECO state in such a way that the magnetic charge degrees of freedom respect to each triangular do not order, which will not give an averaged LRO of spins. Thus, from the point of view of elastic neutron scattering, only diffuse scattering is expected in the absence of sharp magnetic Bragg peaks. In order to clarify the exact ground state of \textit{MgHo}, low temperature neutron diffraction measurement will be necessary.

For $\ast$\textit{ZnHo}, the physics is complicated by the Zn-Ho site-disorder mentioned above.  For a totally disordered TKL system,  the site-disorder  destroys the well-separated Ho kagome layers and forms a disordered 3D pyrochlore-like system where each site has an occupancy of 40\% non-magnetic Zn ions and 60\% magnetic Ho ions. Compared to the Ho pyrochlore lattice, such system will be depleted in the  A-sublattice and stuffed with extra Ho ion in the B-sublattice. In the stuffed spin ice compound Ho$_{2+\delta}$Ti$_{2-\delta}$O$_{7-\delta}$, a ``cluster glass" ground state \cite{stuff} is found. For $\ast$\textit{ZnHo}, the frequency dependence in $\chi_{ac}$ is likely related to a similar glassy behavior.

\subsection{\textit{MgEr} and $\ast$\textit{ZnEr}}
The magnetic properties for \textit{MgEr} were reported in our previous study \cite{TKL}. For $\ast$\textit{ZnEr}, a high temperature Curie-Weiss fit from 100 K to 300 K of  $1/\chi_{dc}$ (Fig. \ref{Fig:Chi} (g)) yields  $\theta_{W}$ = -16.08 K and $\mu_{eff}$ = 9.67 $\mu_{B}$.  These values are similar to that of \textit{MgEr} ($\theta_{W}$ = -14.52 K, $\mu_{eff}$ = 9.45 $\mu_{B}$) and the value of $\mu_{eff}$ is close to the free ion moments of $\mu_{eff}$ = 9.59 $\mu_{B}$ expected for Er$^{3+}$.

For \textit{MgEr}, two  transitions, at 2.1 K and 80 mK are observed. The 80 mK one has been excluded as a LRO transition by our recent neutron scattering experiment \cite{Neutron}, which suggests that no LRO is found in this system down to 50 mK. Meanwhile, the nature of the 2.1 K transition remains a mystery.  It is a robust feature that reproduces exactly, and is therefore unlikely due to a second crystallographic phase in the samples.  The small size of the 2.1 K feature suggests an origin in a phase that is not ``topologically" connected to a classical ordered phase.  Given that Er is $XY$-like in the pyrochlores, we speculate that, if also $XY$-like in the TKL, this phase is a KT vortex unbinding transition.

\begin{figure}[tbp]
	\linespread{1}
	\par
	\begin{center}
		\includegraphics[width= \columnwidth ]{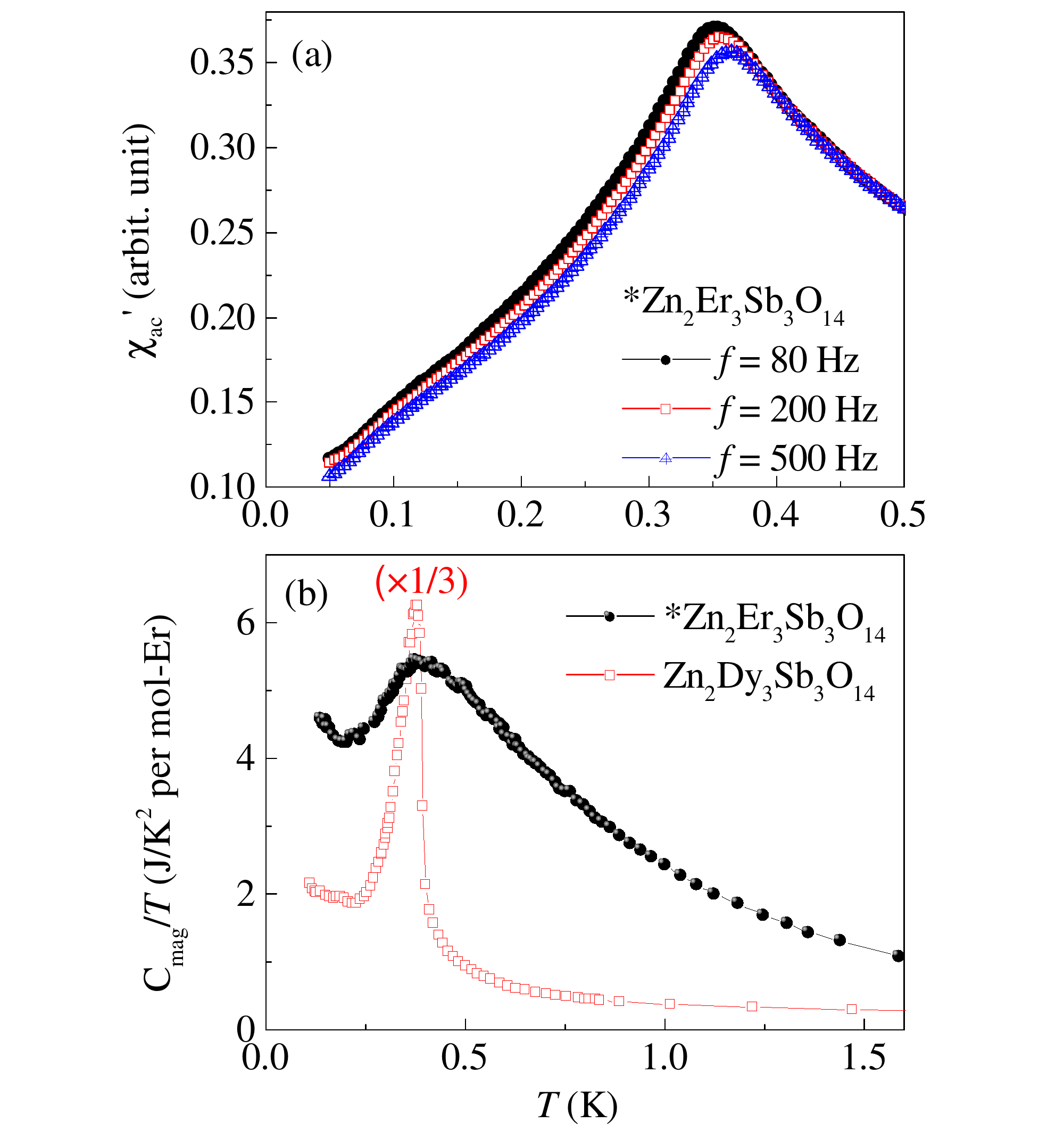}
	\end{center}
	\par
	\caption{\label{Fig:Er}(color online) Temperature dependence of (a) the real part of $\chi_{ac}$ with different ac field frequencies and (b) $C_{mag}/T$ for  $\ast$\textit{ZnEr}. $C_{mag}/T$ for \textit{ZnDy} is plotted for reference.}
\end{figure}

In  $\ast$\textit{ZnEr}, a SRO feature at 0.35 K is observed on both $\chi_{ac}'$ and $C_{mag}(T)$. As shown in Fig. \ref{Fig:Er}, the  $\chi_{ac}'$ peak is broad with an obvious frequency dependence. The peak in $C_{mag}$ is also broad compared with the sharp LRO transition in \textit{ZnDy}. The frequency dependence of $\chi_{ac}'$ and the broadness of the transition from $\chi_{ac}'$  and $C_{mag}$ are characteristic behaviors of a SG system \cite{SG}. Similar to $\ast$\textit{ZnHo},  site disorder exists in $\ast$\textit{ZnEr}. The present situation is akin to (Eu$_x$Sr$_{1-x}$)S \cite{EuSr} and other insulating SG systems where the concentration of magnetic ions is close to the percolation threshold for nearest neighbor interactions.   Therefore, $\ast$\textit{ZnEr} most likely exhibits a SG transition with $T_{SG}$ = 0.35 K.

\subsection{\textit{MgYb} and $\ast$\textit{ZnYb}}
For \textit{MgYb}, a low temperature CW fit from 2 K to 10 K of  $1/\chi_{dc}$ (Fig. \ref{Fig:Chi}(h)) yields  $\theta_{W}$ = -0.45 K and $\mu_{eff}$ = 3.24 $\mu_{B}$.  For $\ast$\textit{ZnYb}, a similar fit gives  a $\theta_{W}$ = -0.39 K and $\mu_{eff}$ = 3.18 $\mu_{B}$. These values for $\mu_{eff}$ are smaller than the free ion moments of $\mu_{eff}$ = 4.54 $\mu_{B}$ expected for Yb$^{3+}$, but similar to those found in the Yb pyrochlores, indicating a planar spin anisotropy at low temperature.  Unlike the small  positive $\theta_{W}$ found in Yb pyrochlores, the values for $\theta_{W}$ for both Yb TKLs are negative.  This implies an enhanced AFM exchange interaction while transforming from the 3D pyrochlore lattice to the 2D TKL. 

\begin{figure}[tbp]
	\linespread{1}
	\par
	\begin{center}
		\includegraphics[width= \columnwidth ]{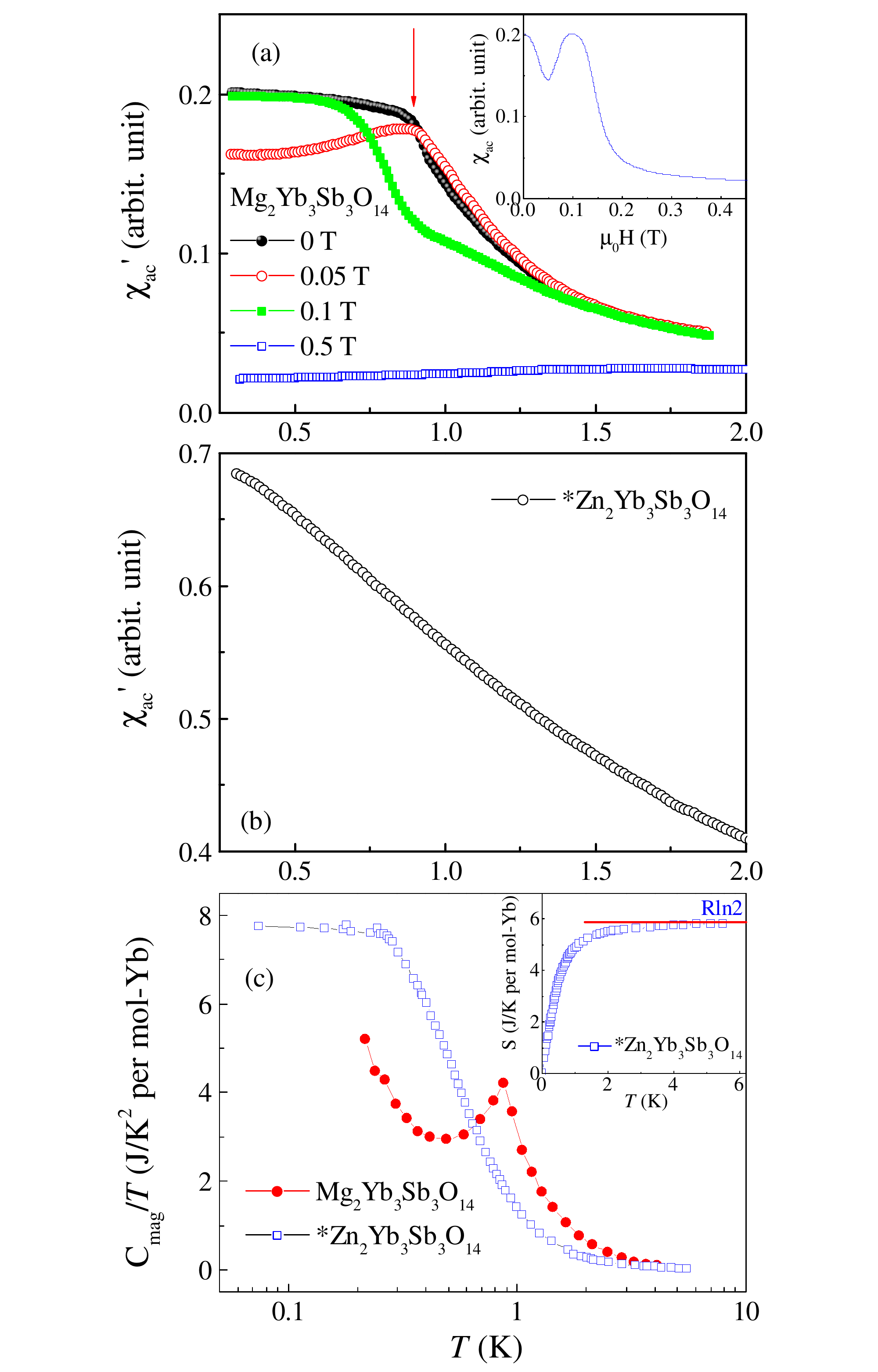}
	\end{center}
	\par
	\caption{\label{Fig:Yb}(color online) (a) Temperature dependence of the real part of $\chi_{ac}$ under different dc field  for \textit{MgYb}. Inset: real part of $\chi_{ac}$ as a function of applied dc field. (b) The real part of the $\chi_{ac}$ for $\ast$\textit{ZnYb}. (c) $C_{mag}/T$ for \textit{MgYb} and $\ast$\textit{ZnYb}. Insert: integrated entropy of $\ast$\textit{ZnYb} assuming $T$-linear behavior of $C_{mag}$ down to zero temperature. }
\end{figure}

For \textit{MgYb},  $\chi_{ac}'$ shows an inflection point at 0.88 K at zero field (Fig. \ref{Fig:Yb}(a)), which is  frequency independent (not shown here). The transition temperature is consistent with a $\lambda$-shape peak in $C_{mag}/T$ (Fig. \ref{Fig:Yb}(c)), suggesting an AFM LRO transition with $T_N$ = 0.88 K. The feature in $\chi_{ac}'$ becomes a well defined peak under a small dc magnetic field of 0.05 T (red curves in Fig. \ref{Fig:Yb}(a)). With even larger dc field, this peak moves to lower temperature and becomes weaker in intensity. The dc field scan of $\chi_{ac}'$ measured at 50 mK is shown in the inset of Fig. \ref{Fig:Yb}(a). Two features are evident from the data: a drop between 0 and 0.05 T and a broad peak around 0.1 T, which drops quickly with even higher field.  Similar behaviors have been observed in a parent pyrochlore antiferromagnet  Yb$_2$Ge$_2$O$_7$. There, the Yb-sublattice possesses an AFM LRO ground state with $T_N$ = 0.62 K. With an applied magnetic field, a double peak feature is  observed in the field scan of  $\chi_{ac}'$, where the first peak around 0.1 T is due to magnetic domains alignment and the second peak at 0.2 T corresponds to a spin-flop transition from the AFM LRO state to the spin polarized state \cite{YbGe}. Similar physics is likely to occur in \textit{MgYb} such that the two features in $\chi_{ac}'$ are due to magnetic domain movement and spin polarization, respectively.

For $\ast$\textit{ZnYb}, paramagnetic behavior is observed in $\chi_{ac}'$ down to the lowest measured temperature of 0.3 K and no LRO is observed in $C_{mag}$ down to 75 mK. Instead,  $C_{mag}/T$ becomes a constant below 0.25 K (Fig. \ref{Fig:Yb}(c)), indicating a $T$-linear behavior for $C_{mag}$. If such  $T$-linear behavior is extended to zero temperature, the integrated entropy from 0 K to 6 K reaches 5.80 J/K per mole-Yb (Fig. \ref{Fig:Yb} (c) inset), which is close to the value of Rln2 = 5.76 J/K expected for an ordered two-level system.  Due to strong site-disorder, some SRO glassy behavior is expected similar to that of $\ast$\textit{ZnEr}. Since no SRO feature is observed in  $\chi_{ac}'$ nor $C_{mag}$, it is possible that the spin freezing process lies below 0.3 K in the $C_{mag}$ $\propto$ $T$ region, which is not detected by $\chi_{ac}'$. On the other hand, if such a possibility is ruled out by further measurements,  the absence of spin freezing and the fully recovered entropy clearly differentiates $\ast$\textit{ZnYb} from a conventional SG system, indicating a single (or very limited number of) micro-state in configuration space at zero temperature. Given the small effective spin-1/2 moments of Yb$^{3+}$, theoretical interpretation of such a state will be interesting even for a system with severe disorder.

\section{Discussion}
\subsection{Spin anisotropies}
In TKLs, if we assume a single ion anisotropy of each RE$^{3+}$ ion similar to that of  their parent pyrochlore lattices, three types of anisotropies are expected. For  Pr$^{3+}$, Nd$^{3+}$, Tb$^{3+}$, Dy$^{3+}$, Ho$^{3+}$, Ising anisotropy is expected, while Er$^{3+}$ and Yb$^{3+}$ should exhibit $XY$ behavior and Gd$^{3+}$ should exhibit Heisenberg-like behavior. In TKL systems, when mediated by exchange/dipolar interactions, these three types of spins introduce different magnetic ground states. The magnetic properties of the 16  A$_2$RE$_3$Sb$_3$O$_{14}$ compounds (A = Mg, Zn; RE = Pr, Nd, Gd, Tb, Dy, Ho, Er, Yb) have been summarized in Table \ref{Tab:2}. Here, aside from the non-magnetic ground state found for Pr-TKLs, we will focus on the Mg branch and discuss their ground states in terms of each type of spin anisotropy.

Among the four compounds Mg$_2$RE$_3$Sb$_3$O$_{14}$ (RE = Nd, Tb, Dy, Ho) with possible Ising anisotropy, \textit{MgNd} have smaller effective moments where the dipole-dipole interaction is small. Although the results from Ref. \cite{NdTKL} seem to suggest an easy-plane ($XY$) spin anisotropy for \textit{MgNd}, the observed all-in-all-out spin structure agrees with the LRO scenario expected for an AFM Ising model. 
The other three Ising spin compounds, Mg$_2$RE$_3$Sb$_3$O$_{14}$ (RE = Tb, Dy, Ho), possess a large moment $\sim$ 10 $\mu_B$. Within this group, the Dy and Ho compounds are most similar due to a small AFM exchange interaction as estimated from the low temperature $\theta_{W}$ ($\sim$ -0.2 K). Since the dipole-dipole interaction has an energy scale $\sim$ 1.3 K, it is tempting to view both systems as pure dipolar ferromagnets. It is possible both compounds possess an ECO ground state, yet apparent difference in $\chi_{ac}$ are observed, suggesting different spin dynamics. This situation of two distinct ground states for the Dy and Ho compounds in TKLs is different from that of pyrochlores where typical spin ice behavior is observed in all Ho$_2$X$_2$O$_{7}$ and Dy$_2$X$_2$O$_{7}$ (X =Ti, Sn, Ge) \cite{HoTi, Hoac, HoGe, Science2001, Nature1999, DySn,  DyGe}.  This interesting contrast suggests the importance of studying the underlying spin dynamics in order to understand precisely how lowering the dimensionality in the TKLs leads to LRO.

In the Mg branch of the TKL family, \textit{MgGd} is the only Heisenberg system due to the half filled f-shell of Gd$^{3+}$. As discussed in Ref. \cite{TKL}, the LRO transitions at 1.65 K is likely due to strong dipole-dipole interactions, which provides an experimental example of the suppression of frustration in a kagome lattice by strong long-range interactions.

The Er$^{3+}$ and Yb$^{3+}$ are both effective spin-1/2 Kramers doublet ions that likely posses $XY$ anisotropy as in their pyrochlore counterparts.  As Yb$^{3+}$ possesses a much smaller moment than Er$^{3+}$, one would expect larger quantum fluctuations which usually perturbs LRO.   In the TKLs, however, LRO is found in \textit{MgYb} instead of \textit{MgEr}. If the $XY$ anisotropy is preserved in both \textit{MgEr} and \textit{MgYb}, our observations imply that the anisotropic exchange that couples to the CEF g-tenser plays an important role. 
It is known from the $XY$ pyrochlore compounds that the detailed balance between anisotropic nearest neighbor exchange interaction, $J_{ex} = (J_{zz}, J_{\pm}, J_{z\pm}, J_{\pm\pm})$, in additional to the strong quantum spin fluctuations of the effective spin-1/2 moment, stabilizes various exotic magnetic ground states \cite{PRLLB}. In Er$_2$Ti$_2$O$_7$, for example, dominant interactions are coplanar type $J_{\pm}$ and $J_{\pm\pm}$ that couples the spin component within the $XY$ plane \cite{ErTiLB}. Accordingly, the magnetic ground state is an AFM state with $XY$-type LRO that is stabilized by quantum fluctuations \cite{ErTiLB, ErTiChampion}. Meanwhile, Yb$_2$Ti$_2$O$_7$ has a major Ising type contribution ($J_{zz}$) to $J_{ex}$ that couples the Ising component of Yb$^{3+}$ moment. Thus, moments in Yb$_2$Ti$_2$O$_7$ tends to behave as Ising spins, which results in a quantum spin ice ground state \cite{YbTi2}. An analogous stabilization is likely to occur in the $XY$ TKLs, which could lead to a complex ground state phase diagram and differences between \textit{MgEr} and \textit{MgYb}.

\subsection{Kramers versus non-Kramers}
We can use another way to categorize the eight TKLs in the Mg branch.  Five of them (RE = Nd, Gd, Dy, Er, Yb) have Kramers ions, whose single ion ground state doublet is restrictively protected by time-reversal symmetry, and are thus degenerated in energy for a mean field of zero. The other three (RE = Pr, Tb, Ho) have non-Kramers ions, whose single ion ground state could also be a doublet but not necessarily degenerate.

In the pyrochlore system, an ``accidental" degeneracy of the non-Kramers doublets is usually found due to protection of a high symmetry point group ($D_{3d}$) at the RE site. Recently, it has been theoretically proposed that in some spin ice like pyrochlore with non-Kramers ions, perturbations such as site-disorder, which acts as local transverse fields, could lift such a degeneracy and possibly lead to different QSL ground states through quantum superpositions of spins \cite{Lucile}. An example is the QSL candidate Pr$_2$Zr$_2$O$_7$, in which a recent inelastic neutron scattering study revealed the lifting of such degeneracy due to a continuous distribution of quenched transverse fields \cite{PrZrCEF}. 

In the TKL system, the rare earth site has reduced its  point group symmetry from $D_{3d}$ to $C_{2h}$. Such a change can be understood crystallographically in terms of the change of local oxygen environment around the RE ion. In the pyrochlore RE$_2$X$_2$O$_7$, one important structural feature is that each RE$^{3+}$ ion is surrounded by eight oxygens with two equivalent RE-O1 bonds lying along the local-[111] axis and six equivalent RE-O2 bonds forming a puckered ring. In a TKL, while the equivalence of two RE-O1 remains unchanged, the six RE-O2 bonds have lost its three-fold rotational symmetry and have been distorted into two sets: four longer RE-O2 bonds and two shorter RE-O3 bonds \cite{TKL}.   In such a case, the accidental degeneracy of the non-Kramers doublet is naturally removed, which splits the doublet into two non-magnetic singlets state with a finite energy difference \cite{CEF}. However, spin-spin interactions which act as local exchange fields can easily mix the two nearby singlets state and recover the magnetic moment.
Starting from the CEF scheme of the parent pyrochlore lattice, if the energy splitting of the two lowest singlet states is comparable to the spin-spin interactions, the additional symmetry reduction can be viewed as a perturbation to the original CEF Hamiltonian where the system remains magnetic with a valid effective spin-1/2 description. One the other hand, if the two lowest singlet states gets too separated in energy, exchange/dipolar interactions will be insufficient to induce magnetism so that a non-magnetic ground state is expected. 

Among the three non-Kramers ion compounds, it is clear that \textit{MgPr} belongs to the second category where a non-magnetic singlet ground state is found.  The other two, \textit{MgTb} and \textit{MgHo}, likely belong to the first category, where the ground state doublets have a finite splitting in energy but remain magnetic. A proper description of the two systems will be  Ising spins under transverse fields on a kagome lattice antiferromagnet. These two compounds can thus be thought of as 2D analogues of  Pr$_2$Zr$_2$O$_7$.
Our classification of the ground states based on Kramers versus non-Kramers ions seems successful: the four TKLs with Kramers ions (except for \textit{MgEr}) exhibit LRO while no LRO is observed for the two TLKs with non-Kramers ions. This result implies that a comprehensive approach considering non-Kramers ions might be needed to explain the absence of LRO in \textit{MgTb} and \textit{MgHo}.

\begin{table*}
	\footnotesize
	\begin{center}
		\caption{ \label{Tab:2} A summary of magnetic properties of  A$_2$RE$_3$Sb$_3$O$_{14}$ (A = Mg, Zn; RE = Pr, Nd, Gd, Tb, Dy, Ho, Er, Yb). For RE =  Nd, Gd, Dy, Ho, Yb compounds, values of $\theta_{W}$ and $\mu_{eff}$ are from low temperature fits of 1/$\chi_{dc}$. For RE = Pr, Tb, Er, values from high temperature fits are used instead because of the nonlinear 1/$\chi_{dc}$ at low temperature due to CEF effects. Therefore, these values from high temperature fits do not necessarily reflect the spin-spin interactions at low temperatures.}
		\renewcommand{\arraystretch}{1.2}%
		\setlength{\tabcolsep}{9pt}
		\begin{tabular}{cccccccccc}		
			\hline 
			\small  &	& Pr & Nd  & Gd & Tb & Dy & Ho & Er & Yb \\ 
			\hline 
			&  $f$ electron (RE$^{3+}$) & 4$f^{2}$ & 4$f^{3}$ & 4$f^{7}$ & 4$f^{8}$ & 4$f^{9}$ & 4$f^{10}$ & 4$f^{11}$ & 4$f^{13}$ \\
			&  Kramers ion (?)  & No & Yes & Yes &  No & Yes & No & Yes & Yes \\
			& Putative anisotropy  & $\sim$ & Ising & Heisenberg & Ising & Ising & Ising & $XY$ & $XY$ \\
				\hline
	       \multirow{4}{*}{A = Mg} 
	          & $\theta_{W}$ (K) & -46.18  & -0.05 & -6.70  & -13.70  & -0.18 & -0.27 & -14.52  & -0.45\\
		      & $\mu_{eff}$ ($\mu_{B}$)  & 3.4 & 2.49 & 8.06 & 9.88 & 10.2 & 10.54 & 9.45 & 3.24\\
		      & Possible Ground state & non-mag. & LRO & LRO & QSL(?) & ECO & ECO(?) KSI(?) & QSL(?) & LRO\\
		      & $T_{N,f,SG}$ (K) & $\sim$ &  0.55 & 1.65 & $\sim$ & 0.37 & 0.4 & 0.08, 2.1 & 0.88 \\
			\hline 
	       \multirow{4}{*}{A = Zn} 
	       & $\theta_{W}$ (K) & -68.43 & -0.11 & -6.85 & -13.41 & -0.72 & -2.49 & -16.08 & -0.39\\
	       & $\mu_{eff}$ ($\mu_{B}$) & 3.61 & 2.28 & 8.09 & 9.86 & 10.2 & 10.22 & 9.67 & 3.18 \\
	       & Possible Ground state & non-mag. & LRO & LRO & QSL(?) & ECO & SG(?) & SG & SG(?)\\
	       & $T_{N,f,SG}$ (K) & $\sim$ & 0.47 & 1.69 & $\sim$ & 0.39 & 0.45 & 0.35 & $\sim$(?)\\		 
			\hline		
		\end{tabular}  \quad
	\end{center}
\end{table*}

\subsection{Chemical pressure effect}
By substituting  the smaller Mg$^{2+}$ ions with the larger Zn$^{2+}$ ions on the non-magnetic A-site in the TKLs, we introduce chemical pressure that enlarges both lattice parameters  $a$ and $c$. Principally, this effect is expected to reduce both the exchange and dipolar interactions. In the pyrochlores, this chemical pressure effect has been proven effective for determining the magnetic ground states. For example, by replacing the Ti site with a smaller Ge ion or a larger Sn ion, the chemical pressure effect selects different magnetic ground states in the pyrochlore systems Yb$_2$X$_2$O$_7$ and Er$_2$X$_2$O$_7$ (X = Ge, Ti, Sn) \cite{YbGe, ErGe}. 

In the TKL system, an obvious result of chemical pressure is the structural change.  As discussed above, while the A/RE site disorder is low in the Mg-branch and Zn-branch with RE ions of larger size, a severe Zn/RE site disorder is present for TKLs with smaller RE ions (RE = Ho, Er, Yb). This type of site disorder destroys the kagome lattice and introduces a random distribution of RE ions with 3D correlation, which will result in a different magnetic ground state.

For TKLs where the site disorder is small (RE = Pr, Nd,  Gd, Tb, Dy), the chemical pressure seems to have little effect on the low temperature magnetism in both branches. Both \textit{MgPr} and \textit{ZnPr} have non-magnetic ground states. We found LRO in the Nd, Gd and Dy compounds in both Mg and Zn branches with similar ordering temperatures (Tab. \ref{Tab:2}), consistent with the small lattice constant differences. No LRO is observed in  both \textit{MgTb} and \textit{ZnTb}. It is noteworthy that some subtle differences have been observed. For example, while \textit{MgTb} shows a broad SRO-like feature in $\chi_{ac}'$ at 400 mK, no such feature is seen in \textit{ZnTb}. In Fig. \ref{Fig:Tb}(c), $C_{mag}/T$ also shows some difference above 400 mK between the two. 
Another example is that for \textit{MgNd}, the zero field $\chi_{ac}'$ seems indicate a two-step transition, while for \textit{ZnNd}, a small magnetic dc field is required to separate them. Also, for \textit{ZnDy}, an extra increase of $C_{mag}$ is observed below the ECO transition, which is absence in that of \textit{MgDy}. More work is needed to understand the differences between these systems.

\subsection{Future directions}
As mentioned in the introduction, the unique structure and rich spin types of the TKLs provide us a platform to realize exotic kagome-based physics. Our $\chi_{ac}$ and $C_{mag}$ measurements have revealed some basic magnetic behaviors for the sixteen TKL members, which will provide a guidance for further studies.

Future CEF excitation measurements will be important for determining the CEF levels, g-tensors, and therefore to confirm the spin anisotropies for TKL members. Since no LRO is observed down to 50 mK in \textit{MgTb}, \textit{ZnTb}, \textit{MgEr}, and \textit{ZnYb}, these four TKL systems are promising candidates for hosting QSL states. Other experiments, including neutron scattering, $\mu$SR, and NMR, will be helpful to identify the nature of their ground states as well as the spin correlations. Since QSL states are characterized by the presence or absence of a gap in the anyon excitation spectrum, it is important to determine if any of the QSL candidate TKLs possess such a gap. For Nd-KLs, the nature of the two-step order needs to be addressed.  Measurements under weak magnetic fields are needed to provide insight into this question.  Interpreting such measurements will be difficult, given the present polycrystalline samples, but coarse-grained behavior can be studied. For Gd-TKLs, our theoretical investigation based on the Luttinger-Tisza method predicts a 120 $^\circ$ spin structure. Confirmations are needed by other direct measurements. For \textit{MgDy}, neutron scattering experiments based on a powder sample with 4-6\% site disorder have identified an ECO with average spin LRO ground state. We expect a similar ground state in \textit{ZnDy}. A question then would be whether or not perfect samples (lacking any site disorder) give rise to true LRO of the Dy$^{3+}$ spins, as predicted by the theory \cite{ECO1,ECO2}. For \textit{MgHo}, we have identified a SRO state, but whether it is a  KSI state or an ECO state with quantum fluctuation needs to be determined.  For \textit{MgEr}, it remains unclear what is the nature of the 80 mK peak in $\chi_{ac}'$ and whether the 2.1 K transition in $C_{mag}$ is indeed a KT transition.  It is also not clear what is the exact LRO state for \textit{MgYb}. Future studies are needed to answer all these questions.

On the theoretical side, due to the uniqueness of the tripod-like local Ising/$XY$ anisotropy on a kagome lattice, there are limited theoretical works at the moment that suitably describe the TKL system. Our previous theoretical investigation using a Luttinger-Tisza type theory provides a first universal mean-field level description of the TKL \cite{TKL}. Aside from this, there are few existing theoretical studies that can be directly adopted. For Heisenberg spins, Moessner et al. considered classical dipoles on a 2D kagome lattice and calculated a phase diagram by scaling dipolar and exchange interactions \cite{Dipoles}. This model might be a good starting point for the Gd-TKLs. For Ising spins, models with TKL-like canted Ising spins have predicted ECO followed by a spin LRO transition on a kagome lattice \cite{ECO1,ECO1}, which seems to successfully explain the experimental observation in \textit{MgDy}. However, obvious disagreements exist for Tb and Ho-TKLs. For $XY$ spins, there are even few (if any) theoretical studies since the situation of three distinct local $XY$ planes has most likely not been previously considered before the realization of the TKL. Therefore, the TKLs offer an unexplored realm of theory. We hope our results will stimulate more theoretical studies on these exciting compounds.\\

\begin{acknowledgments}
Z. L. Dun and H. D. Zhou thank the support of NSF-DMR-1350002. A. P. Ramirez was supported by NSF-DMR 1534741.  Y. X. Wang acknowledge the support of the National Natural Science Foundation of China (Grant
No. 11275012). K. Li acknowledges the support of NSAF (Grant No. U1530402) and NSFC (Grant No. 21501162).  The work at  NHMFL is supported by NSF-DMR-1157490 and State of Florida and the DOE and by the additional funding from NHMFL User Collaboration. 
\end{acknowledgments}


\begin{thebibliography}{99}
\bibitem{note} The refinements of the site disordered TKLs are not as good as those of the site ordered TKLs. The refinements do not give stable oxygen positions. And there are tiny peaks (less than 1\% in intensity of the strongest peak) that can not be indexed by the TKL structure. Though the Zn/RE site order model catches the main feature of the XRD pattern, it is possible that their structure could be more accurately described by a different model.
\bibitem{Neutron} Inelastic neutron scattering experiment shows intense diffuse scattering persisting down to $T$ $\sim$ 60 mK. The data will be published elsewhere.
\end{thebibliography}
\end{document}